\documentclass[usenatbib]{mn2e}
\usepackage{color} 
\usepackage{graphics}
\usepackage{epsfig}
\usepackage{natbib}

\begin{document}

\title[Galaxy centers with young and unusually massive stars] 
{A study of the central stellar populations of galaxies in SDSS-IV MaNGA:   
identification of a sub-sample with unusually young and massive stars}                             

\author [G.Kauffmann] {Guinevere Kauffmann$^1$\thanks{E-mail: gamk@mpa-garching.mpg.de}\\
$^1$Max-Planck Institut f\"{u}r Astrophysik, 85741 Garching, Germany}

\maketitle

\begin{abstract} 
This paper describes a search for galaxy centers with clear
indications of unusual stellar populations  
with an initial mass function flatter than
Salpeter at high stellar masses. 
Out of a sample of 668 face-on galaxies with stellar masses in the range
$10^{10}- 10^{11} M_{\odot}$, I identify 15 galaxies with young to
intermediate age central stellar populations with unusual stellar population
gradients in the inner regions of the galaxy. In these
galaxies, the 4000 \AA\ break is either flat or rising towards the center
of the galaxy, indicating that the central regions host evolved stars,
but the H$\alpha$ equivalent width also rises steeply in the central regions.
The ionization parameter [OIII]/[OII] is typically low in these  galactic
centers, indicating that ionizing sources are stellar rather than AGN. Wolf
Rayet features characteristic of hot young stars are often found in the
spectra and these also get progressively stronger at smaller galactocentric
radii.
These outliers are compared to a control sample of galaxies of similar mass
with young inner stellar populations, but where the gradients in H$\alpha$
equivalent width and 4000 \AA\ break follow each other more closely. 
The outliers exhibit central Wolf Rayet red bump excesses much more frequently, they
have higher central stellar and ionized gas
metallicities, and they are also more frequently detected at 20 cm
radio wavelengths. I highlight one outlier where the
ionized gas is clearly being strongly perturbed and blown out either by
massive stars after they explode as supernovae, or by energy injection from
matter falling onto a black hole.
\end {abstract}

\section{Introduction}

Studies of the demographics of large samples of galaxies with active galactic
nuclei in the local Universe  have demonstrated that most present-day accretion
occurs onto black holes with masses less than $10^8 M_{\odot}$ that reside in
moderately massive galaxies ($10^{10}-10^{11} M_{\odot}$) with high central
stellar surface mass densities and young stellar populations (Kauffmann et al 2003).
The volume-averaged ratio of star formation to black hole accretion in
these systems is $\sim 1000$ (Heckman et al 2004), a value that is remarkably close to the the
observed ratio of stellar mass to black hole mass in nearby galaxy bulges (Ferrarese \& Merritt 2000,
Gebhardt et al 2000).
This implies that black hole growth and bulge growth through star formation are
tightly coupled 

Up to recently, combined studies of black hole and bulge growth in complete
samples of galaxies have been confined to data from optical single-fibre
spectroscopic surveys at low redshifts (e.g. Kauffmann et al 2003; Kewley et
al 2006), or joint analysis of X-ray imaging data and optical/near-infrared surveys of
galaxies at higher redshifts (e.g. Hickox et al 2009). With the advent of integral field unit (IFU)
spectroscopic surveys that  measure spectra for hundreds of locations  within each
galaxy, it has become possible to study  the spatial distribution and nature of the
young stars and their surrounding ionized gas within growing bulges, thereby
gaining detailed insight into the observed coupling between star formation and
black hole formation in these systems.

It has long been recognized that high mass bulges with stellar
masses greater than $\sim 3 \times 10^{10} M_{\odot}$   are similar to elliptical
galaxies in having predominantly old and metal-rich stellar populations and in
falling on much the same scaling relations between age/metallicity and bulge
velocity dispersion as the ellipticals (Balcells \& Peletier 1994; Terndrup et
al 1994; De Jong 1996; Thomas et al 2005). Low mass bulges 
($M_* \sim 10^9-10^{10} M_{\odot}$)  exhibit much more
scatter in both age and metallicity.  Resolved spectroscopic studies of small
bulge samples have shown that they exhbit a wide range of age and metallicity
gradients, from negative to positive, indicating a diversity of formation
histories and mechanisms (MacArthur et al 2009).

More than a half of galaxy bulges in late type galaxies with stellar masses in
the range $10^8-10^{11} M_{\odot}$ contain nuclear star clusters (NSC) at their
very centers. These are luminous and compact sources of excess light above the
inward extrapolation of the host galaxy's surface brightness profile that are
found  on scales of 50pc or less (see Neumayer et al 2020 for a recent review).
The NSC  at our own Galactic Center, first identified by Becklin \& Neugebauer
(1968) is the most well-studied of these systems. It appears to be intrinsically
elliptical and flattened along the Galactic plane, with an axis ratio $b/a=0.71$
(Sch\"odel et al 2014).  It is also rotating in the same direction as the
Galactic disk (Genzel et al  1996).  The central 0.5 pc region of the Milky Way
NSC hosts 200 massive and young Wolf-Rayet and O- and B-type stars that are
partitioned between two rotating disk-like structures (Genzel et al 2003; Levin
\& Beloborodov 2003).  Because of the young ages (6-8 Myr) inferred for these
stars, they are believed to have formed in situ within very dense, accreted gas.
The initial mass function (IMF) of the stars inferred from their K-band
luminosity function is inferred to be considerably flatter than Salpeter
(Paumard et al 2006), though the exact form of the derived IMF appears to be
quite sensitive to location within these structures (Lu et al 2018).

In external galaxies, all but the closest NSCs are barely resolved even with the
Hubble Space Telescope. In general, the  stellar populations of most nuclear
star clusters are characterized by a dominant old stellar population plus a
significant population of young stars. The NSCs also tend to be more metal rich
than their surrounding host galaxy.

The formation of stellar mass black holes in NSCs and their possible evolution
to intermediate masses through runaway collsions has been a topic of theoretical
speculation for many years (see for example Rees 1978; Portegies Zwart  \&
MacMillan 2002), but rather few observational constraints currently exist.  One key
unknown is the nature star formation in the extreme environments typical of
galaxy centers. Our own Milky Way is believed to have had a relatively quiescent
formation history.  Questions such as ``do most stars form with the same initial
mass function in bulges  as in galactic disks", ``what is the fraction of stars
formed in bound systems rather than in isolation", and ``how does a very dense
stellar and gaseous environment influence stellar evolution" remain largely
unanswered for the galaxy population as a whole.

In this paper, I analyze the central stellar populations of a sample of 1000
galaxies with stellar masses in the range $10^{10}-10^{11} M_{\odot}$ with
face-on orientation and with IFU data from the Mapping Nearby Galaxies at APO
(MaNGA) survey. Face-on galaxies are chosen so as to minimize the effects of  
dust attenuation and allow the stars in circumnuclear disk-like structures to be detected
more easily (see Neumayer et al 2020). The goal of this paper is to identify
bulges with central stellar populations  indicative of ``non-standard"
phenomena, such as an excess of high mass stars compared to typical disk stellar
populations.

Stellar population synthesis models are often used as a way to constrain
parameters such as the age, metallicity and IMF of a galaxy. Model-fitting is
always subject to considerable systematic uncertainties that arise from a
variety of different factors (see Conroy et al 2009 for a review). The approach that
is taken here is empirical. Section 2 describes the observations and the
sample selection.  Section 3 examines scaling relations between three
spectroscopic indicators of stellar age: the 4000 \AA\ break strength, the
Balmer absorption line index strength H$\delta_A$ and the 
H$\alpha$ line equivalent width corrected for dust attenuation. 
Strong correlations  are found between these
quantities in the outer galaxy. The majority of galaxies exhibit inner stellar
population scalings that lie on the same relations as disks. In a minority of
galactic bulge regions, there are strong deviations from the disk scalings.

In section 4, bulges with strongly deviating {\em central} stellar
populations are identified  by examining inner gradients for three stellar population
indicators and how they correlate with each other. A bulge with a deviating
central stellar population is defined as one where the correlations between the
stellar indicator gradients differ strongly from those defined by the  majority
of the sample. Section 5 presents a detailed analysis of the stellar and ionized gas
properties of the sample of bulges with unusual centers, comparing and
contrasting the results with those found for control samples of bulges with
``normal" central stellar indicator gradients. Section 6 presents some first
insights from stellar population synthesis modelling and discusses possible
ways to investigate linkages with 
black hole formation and accretion. Finally, the main results from the paper
are summarized in Section 7.

\section {The parent galaxy sample}

The sample of galaxies is drawn from the 15th data release (DR15) of the Sloan
Digital Sky Survey's  MaNGA project (Bundy et al 2015), which is part of the
Sloan Digital Sky Survey IV programme (SDSS-IV; Blanton et al. 2017). The final
MaNGA sample will consist of approximately 10,000 galaxies with redshifts
$0.01 < z < 0.18$, spanning a wide range in galaxy morphologies and selected to be
have an approximately flat distribution in terms of log(stellar mass) (Yan et al. 2016;
Wake et al. 2017). Observations are taken using the BOSS spectrograph (Smee et
al. 2013) on the 2.5 m Sloan telescope at Apache Point Observatory (Gunn et al.
2006). The spectra have a wavelength range of 3600-10000 \AA\ and a spectral
resolution of R $\sim$ 2000 (Drory et al. 2015).  MaNGA's individual IFUs
consist of hexagonal fibre bundles containing 19-127 optical fibres (each with
diameter 2 arcsec), with a threepoint dithering pattern employed during
observations in order to fully sample the targetted field of view (Drory et al.
2015; Law et al. 2015). Observations are reduced through the MaNGA Data
Reduction Pipeline (DRP; Law et al. 2016; Yan et al. 2016), which
flux-calibrates and sky-subtracts spectra before drizzling observations onto
spaxels of width 0.5 $\times$ 0.5 arcsec. Flux calibration is performed using
standard stars observed with fibre bundles of 7 fibres apiece. The final
reconstructed datacubes have a point-spread function (PSF) full-width at
half-maximum (FWHM) of approximately 2.5 arcsec (Law et al. 2015).

DR15 includes data for 4,621 unique galaxies. The available data consists of the
raw data from the first three years of the survey, the intermediate/final data
reduction pipeline (DRP; Law et al 2016) products, and the first release of
derived data products from the data analysis pipeline (DAP;  Westfall et al.
2019; Belfiore et al. 2019).  These derived data products include maps of
emission line fluxes, gas and stellar kinematics, and stellar population
properties.  The DAP also includes model fits to the stellar continuum using the
pPXF fitting routine (Cappellari \& Emsellem 2004), which employs a stellar-template
library constructed by hierarchically-clustering the MILES stellar library
(Vazdekis et al 2016) into a set of 42 composite spectra, termed the MILESHC.

Once the stellar-continuum fit has been performed, the DAP analyzes the
emission-lines by subtracting the best-fitting continuum model from the data.
In this paper, the unbinned per-spaxel emission line and spectral line index
measurements available from the MAPS files are used where possible.  The MAPS
files are the primary output file from the DAP and provide 2D ``maps'' (i.e.,
images) of DAP measured properties.  Model LOGCUBE files, which provide the
binned spectra and the best-fitting model spectrum for each spectrum that was
successfully fit, are used  to compute additional spectral parameters such as
spectral ``bumps" indicating the presence of Wolf-Rayet stars (see section 5 for
more details).

Stellar masses and structural parameters such as position angles, ellipticities,
and half-light radii for all galaxies in the MaNGA sample are available from the
MaNGA {\em drpall} file; in all cases, the structural parameters are obtained
from the SDSS elliptical Petrosian apertures. All galaxies with ellipticity
parameter $b/a$ greater than 0.7, half-light radii greater than 5 arcsecond and
stellar masses in the range $10^{10} < \log  M_* < 10^{11} M_{\odot}$ are
selected. The stellar mass cut is adopted to confine this analysis to galaxies
that are similar in mass to our Milky Way. As discussed in section 1, the cut on
galaxies with near face-on inclination helps to minimize dust obscuration of
young stellar populations near the centers of galaxies. The cuts yield a sample
of 668 galaxies with flags that indicate good quality data.

\begin{figure*}
\includegraphics[width=141mm]{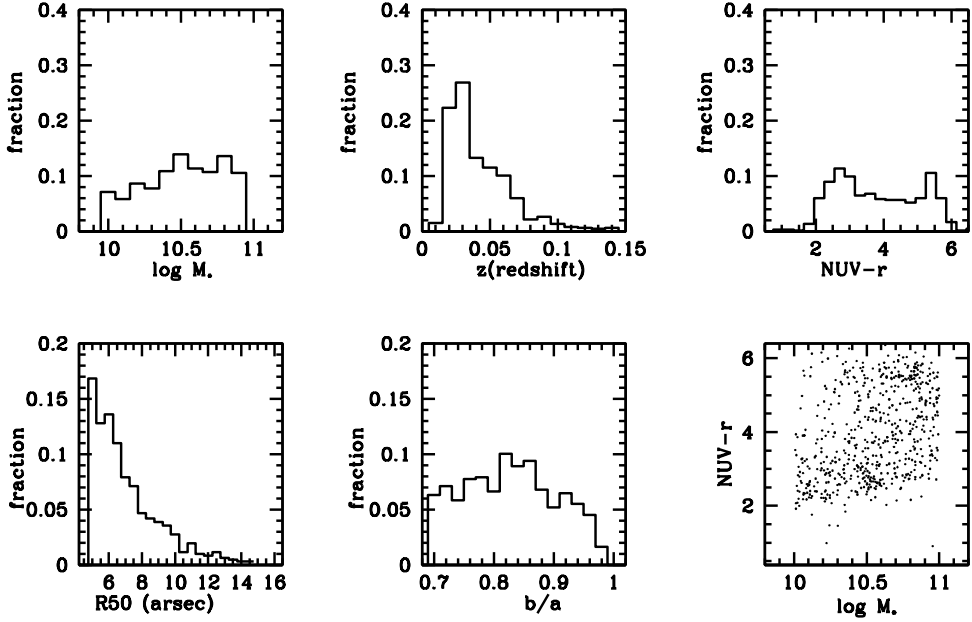}
\caption{ The global properties of the parent galaxy sample.  Histograms
of stellar mass, redshift, global NUV-r colour, 
r-band half light radius R$_{50}$ and ellipticity
parameter b/a are presented in the first 5 panels. The 6th panel
shows a scatter plot of NUV-r colour versus stellar mass.
\label{models}}
\end{figure*}

Figure 1 presents the basic parameters of the galaxies in this sample. Histograms
of stellar mass, redshift, global NUV-r colour (the NUV photometry is taken from
imaging data by the GALEX satellite and is included as part of the NASA-Sloan Atlas
\footnote{http://nsatlas.org}), r-band half light radius R$_{50}$ and ellipticity
parameter $b/a$ are presented. As can be seen, the galaxies are distributed
fairly evenly in stellar mass, colour and ellipticity within the selection 
boundaries. Almost all  (93\%) are located at redshifts less than 0.07 and the typical
radius of the half-light region is 7-8 arcsec.

\section {Comparison of radially averaged inner and outer stellar populations}
The analysis begins by examining the correlations between 3 different indicators
of the stellar age of a galaxy for the full sample
of 668 galaxies. These are  \begin {enumerate} \item The narrow
definition of the  4000 \AA\ break, defined as the ratio of the average flux
densities $F_{\nu}$ in the bands 3850-3950 and 4000-4100 \AA\ (Balogh et al.
1999).  This index is denoted as as D$_n$(4000) throughout this paper.  \item
The H$\delta_A$ index (Worthey \& Ottaviani 1997) with central index bandpass
4083.5 - 4122.25 \AA.  \item The $H\alpha$ emission line equivalent width,
denoted EQW H$\alpha$.
\end {enumerate}
These indicators all probe the age of the stellar population,
but are sensitive to light from stars of different masses, and scaling
relations between them should be sensitive probes of the stellar initial
mass function (IMF).

When calculating the $H\alpha$ equivalent width, the $H\alpha$ line flux is
corrected for dust attenuation using the measured Balmer decrement using the
formula $A_V=1.9655 R_V \log(H\alpha/H\beta/2.87)$, where $R_V=3.1$ and  the
Calzetti (2001) attenuation curve have been adopted.  The stellar continuum
measurements are not corrected for dust attenuation.  This is best done in the
context of full-spectrum fitting using stellar population synthesis models and
will be the subject of future work. This paper focuses on relative trends in
directly measured quantities as a function of position within the galaxy.

For each galaxy, radially averaged values of each stellar indicator are
evaluated in bins of R/R$_{50}$ where R$_{50}$ is the half-light radius. The bins range
from 0.1 R/R$_{50}$ to 1.5 R$_{50}$ in steps of 0.1 R/R$_{50}$. Radially averaging
serves to damp fluctuations caused by individual HII regions and results in
tighter correlations than those produced by individual spaxel measurements.

Figure 2 compares correlations between the three indicators for stellar
populations in the inner galaxy (defined as $R < 0.5$ R$_{50}$ and plotted as red
points) with those in the outer galaxy (defined as 0.5 R$_{50}$ $< R < 1.5$ R$_{50}$ and
plotted as black points). Results are shown separately for two mass
ranges: $10 < \log M_* < 10.5 M_{\odot}$ (left panels) and 
$10.5 < \log M_* < 11 M_{\odot}$ (right
panels) to investigate whether there are any systematic trends
with the mass of the galaxy.

\begin{figure}
\includegraphics[width=92mm]{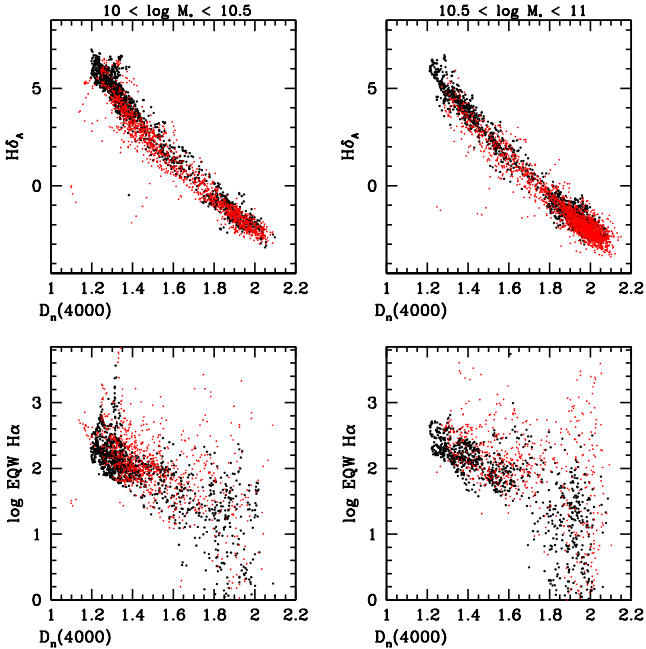}
\caption{Correlations between the three indicators for stellar
populations (D$_n$(4000), H$\delta_A$ and EQW H$\alpha$) are shown for
radially averaged measurements  in the inner galaxy (defined as $R<0.5$ R$_{50}$ and plotted as red
points) and in the outer galaxy (defined as 0.5 R$_{50}$ $< R < 1.5$ R$_{50}$ and
plotted as black points). Results are shown separately for two mass
ranges: $10 < \log M_* < 10.5$ (left panels) and $10.5 < \log M_* < 11$ (right
panels). All inner galaxy spaxels that have emission line ratios
consistent with AGN ionization have been removed (see text).  
\label{models}}
\end{figure}

Radially averaged stellar populations occupy a very tight sequence in the plane
of H$\delta_A$ versus D$_n$(4000). Inner and outer stellar populations largely
overlap each other with  a hint of greater scatter for the inner galaxy stellar
populations. The red points are more evenly spread along the H$\delta_A$ versus
D$_n$(4000) locus for lower mass galaxies than for higher mass galaxies,
indicating a shift towards older ages in the bulges of higher mass galaxies.
This is in accord with previous findings in the literature, 
as outlined in Section 1.  In
particular, the H$\alpha$ EQW distribution in the inner galaxy exhibits a 10\%
tail to values in excess of a few hundred; this tail is not seen in the outer
galaxy. This will form the subject of more detailed investigation later in this
paper.

The relation between H$\alpha$ equivalent width and D$_n$(4000) shown in the
bottom panel of Figure 2 exhibits considerably more scatter, particularly at
large values of D$_n$(4000). One possible source of scatter in the central
regions of galaxies not related to stars, is ionization from AGN.
In order to minimize this effect, in the bottom
two panels, we have made a cut in the [NII]$\lambda$6584/H$\alpha$ versus
[OIII]$\lambda$5007/H$\beta$ BPT (Baldwin, Philips \& Terlevich 1981) 
nebular emission line diagram, excluding all spaxels with emission line
ratios that satisfy the  AGN cut given in 
Kauffmann et al (2003): \begin{equation} \log([{\rm OIII}]/H\beta) >
0.61/(\log([{\rm NII}]/\rm{H}\alpha)-0.05)+1.3. \end{equation}
This cut is only applied to spaxels from the inner regions
of the galaxies and it removes 24\% of the inner spaxels from
galaxies in the low mass bin and 31\% of spaxels from galaxies
in the high mass bin.  Even after
exclusion of central spaxels that may be contaminated by AGN emission,  we see a
clear shift in the red points to higher values of H$\alpha$ equivalent width at
fixed D$_n$(4000). This is the first suggestion that the stellar populations
in the central regions of galaxies are systematically different to those
in the outer regions.

The results from Figure 2 are quantified in more detail in Figure 3, where solid
red and black lines show the median relations between H$\delta_A$ and
D$_n$(4000) and H$\alpha$ equivalent width and D$_n$(4000) for outer and inner
stellar populations, respectively, while dashed and dotted lines show the  10th
and 90th percentiles of the distribution of H$\delta_A$  and  H$\alpha$ EQW at a
fixed value of D$_n$(4000). As in the previous figure,
all inner galaxy spaxels that have emission line ratios
consistent with AGN ionization have been removed. To make this figure, the radially averaged
measurements are arranged in ascending order of D$_n$(4000) and the median and
percentiles are computed for bins containing 50 measurements. The up and down
fluctuations in the lines should thus be regarded as a measure of the Poisson
error in the evaluation of these quantities.

\begin{figure}
\includegraphics[width=92mm]{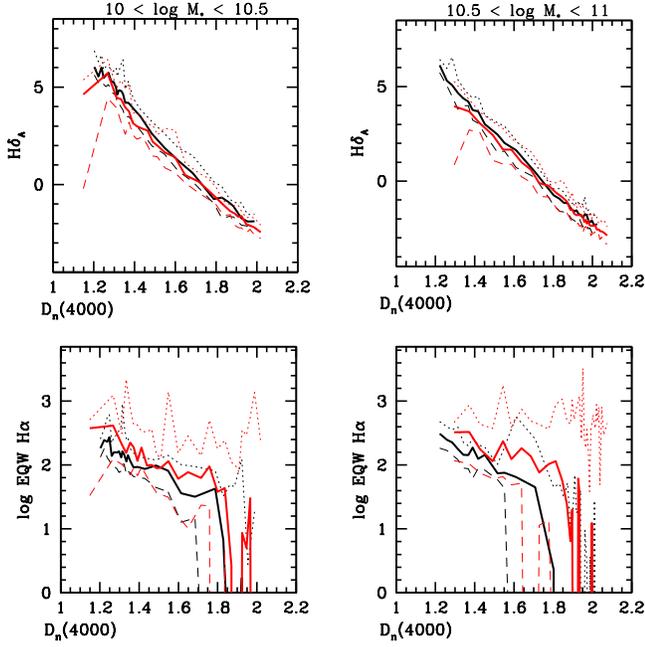}
\caption{ Red and black lines show the median relations between H$\delta_A$ and
D$_n$(4000) and H$\alpha$ equivalent width and D$_n$(4000) for outer and inner
stellar populations, respectively, while dashed and dotted lines show the  10th
and 90th percentiles of the distribution of H$\delta_A$  and  H$\alpha$ EQW at a
fixed value of D$_n$(4000). As in Figure 2, the results are shown in two 
stellar mass bins. 
\label{models}}
\end{figure}

As can be seen, there is a small, but significant shift in the median
H$\delta_A$ towards lower values at fixed D$_n$(4000) for inner stellar
populations compared to outer ones, and a larger shift in the median EQW
H$\alpha$ towards high values for inner stellar populations.  There is also a
significant outlying population of bulges that have radially-averaged EQW
H$\alpha$ values greater than a few hundred.  Such large  H$\alpha$  EQW values
are largely absent in the outer stellar populations of galaxies. These outlier
bulges will form the subject of a more detailed investigation in Section 4 of
the paper.

\section {Inner stellar population gradients}

This section examines radial trends in the three age indicators, H$\delta_A$,
D$_n$(4000) and EQW H$\alpha$, and the degree to which these correlate with each
other. Note that no cuts to remove AGN-contaminated spaxels are made in this 
section.

For each galaxy, the spaxel measurements are divided into 4 radial
ranges: 0-0.3 R$_{50}$, 0.3-0.6 R$_{50}$, 0.6-0.9 R$_{50}$ and 0.9-1.2 R$_{50}$.
The slope is determined by linear least squares regression. The index difference
is denoted $\Delta$ and is defined as the  difference in the best fit index
value at the two endpoints of the radial range: for example,  $\Delta$D$_n$(4000)
(0.3-0.6 R$_{50}$) is the difference in the linear fit to  4000 \AA\
break strength versus R/R$_{50}$ evaluated at 0.3 R$_{50}$ and at 0.6 R$_{50}$.

Figure 4  shows three index differences for 4 different radial ranges as a
function of the NUV-r colour of each galaxy. Results for galaxies in the stellar
mass range $10 < \log M_* < 10.5$ $M_{\odot}$ are shown in the first part of  Figure 4 and for galaxies in the
mass range $10.5 < \log M_* < 11$ $M_{\odot}$ in the second part.  As can be seen, gradients in the
stellar age indicators are strongest in the innermost regions of blue galaxies with
NUV-r$<4.5$
and become progressively weaker at larger radial distances. The majority of blue
galaxies  have negative  D$_n$(4000) gradients   and positive  $H\delta_A$ 
gradients in their inner regions, indicating that stellar populations become progressively 
older (higher D$_n$(4000) and lower $H\delta_A$) 
towards the centres of most galaxies. The typical H$\alpha$ EQW profile is
relatively flat in blue galaxies with NUV-r$<4.5$, but increases towards the central regions in
red galaxies with NUV-r$>4.5$, which likely indicates a greater contribution from AGN rather than
from stellar ionization sources in red systems.

\begin{figure}
\includegraphics[width=91mm]{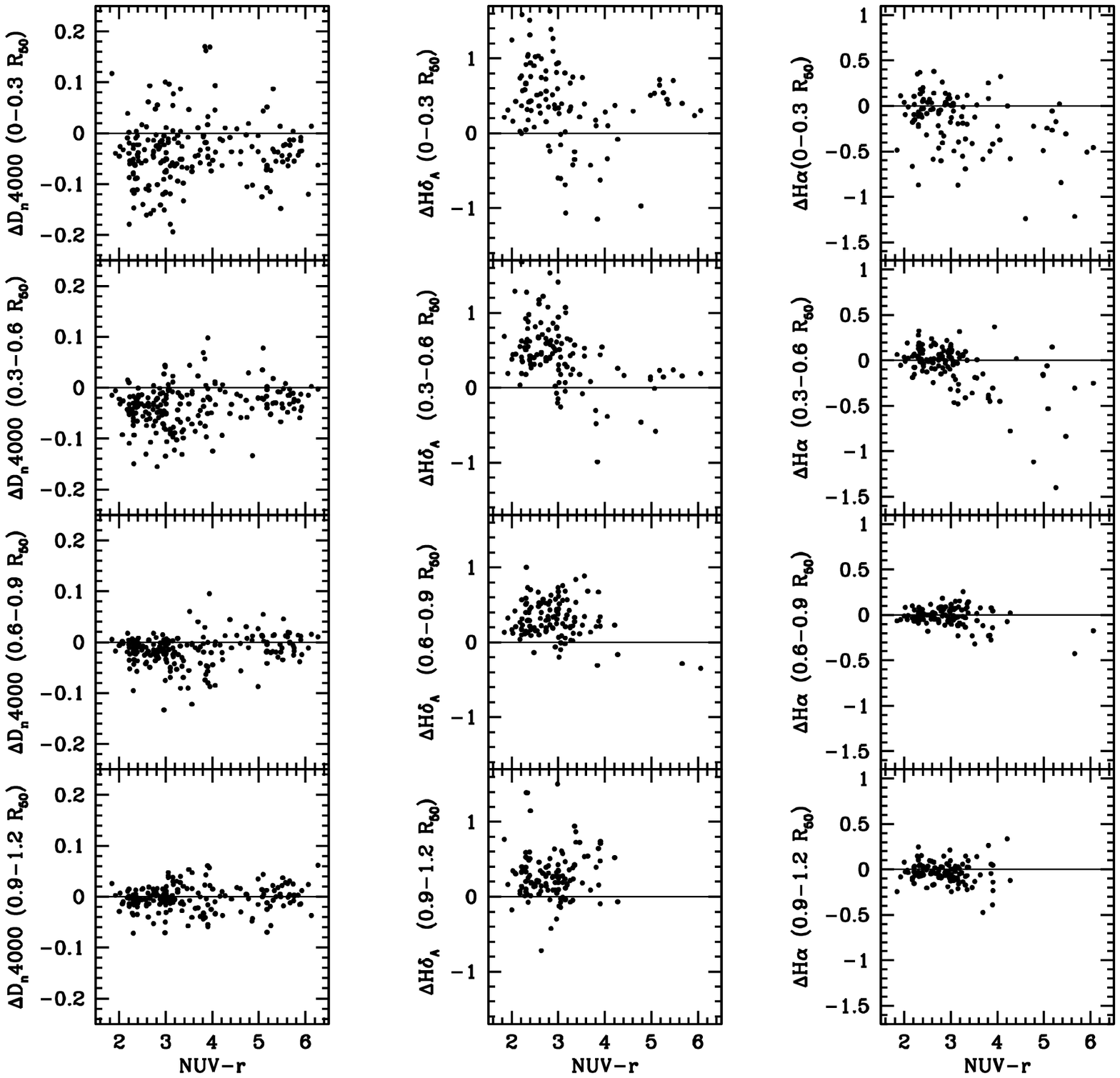}
\caption{The index differences $\Delta$ D$_n$(4000), $\Delta$H$\delta_A$ and
 $\Delta\log$ EQW(H$\alpha$) are plotted as a function of the NUV-r colour of the
galaxy. Results are shown in three rows for 3 different radial ranges for galaxies with 
stellar masses in the range $10 < \log M_* < 10.5 M_{\odot}$}   
\includegraphics[width=91mm]{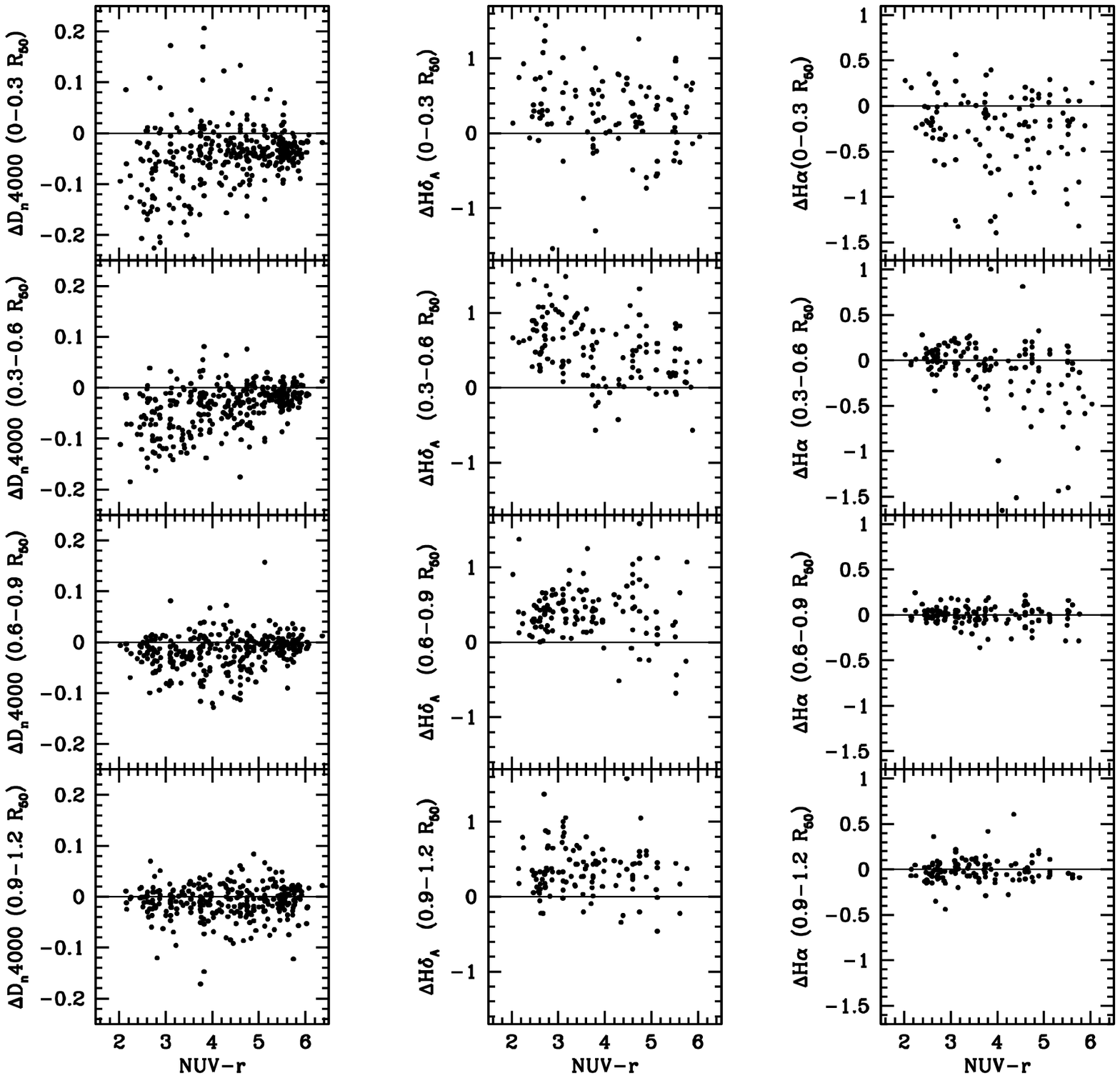}
\contcaption{ Same as the previous sub-figure, except for galaxies with stellar masses in the 
range $10.5 < \log M_* < 11 M_{\odot}$}   
\end{figure}

Results for the low mass and high mass subsamples shown in Figures 4  are
qualitatively similar. One difference is that at intermediate  
NUV-r colours of $\sim 3.5-4.5$, there are around a dozen galaxies
in the lower mass sample  where the
D$_n$(4000) profile is flat in the 0-0.3 R$_{50}$ radial bin, but where
H$\delta_A$ is rising towards the central regions of the galaxy.  These are
likely ``post-starburst" bulges, where there has been a significant central
burst of star formation a few hundred million years to 1 Gyr in the past.  Hot O
and B-type stars have already evolved off the main sequence in these bulges, but
there is still a significant contribution from A to F-type stars that results in
strong Balmer-line absorption in their centres. In the high stellar mass bin,
the D$_n$(4000) and H$\delta_A$ profiles in the central radial bin appear to
track each other in almost all galaxies irrespective of colour.

Figure 5 examines correlations between the index differences for the central
0-0.3 R$_{50}$ radial bin. The reader is reminded that the focus of this paper
is the question of whether there is a population of bulges where there is evidence
for an excess population of young massive stars similar to that
claimed for the central  region of our Milky Way. The presence of young,
massive stars is probed directly by H$\alpha$ line emission, so it is intriguing
that a  population of apparent outliers is found in the plane of $\Delta$(log
EQW H$\alpha$)  versus $\Delta$(D$_n$(4000)) consisting of objects with very
negative $\Delta$(log EQW H$\alpha$) indicative of steeply peaked central
H$\alpha$ emission, but small $\Delta$(D$_n$(4000)) indicating weak variations
in the age of the older stellar populations in the central regions of these
galaxies. A cut \begin{equation} \Delta(\log {\rm EQW
H}\alpha)<(1.0/(-0.222))\Delta({\rm D}_n4000))+0.026-0.6 \end{equation} indicated by a
red line in the middle panels of Figure 5 is used to pull out these systems,
which are plotted as magenta points. As can be seen, these systems do not stand
out in the plane of $\Delta$H$\delta_A$ versus $\Delta$D$_n$(4000), indicating
that the underlying older stellar population follows close-to-normal relations in
these galaxies.

\begin{figure*}
\includegraphics[width=141mm]{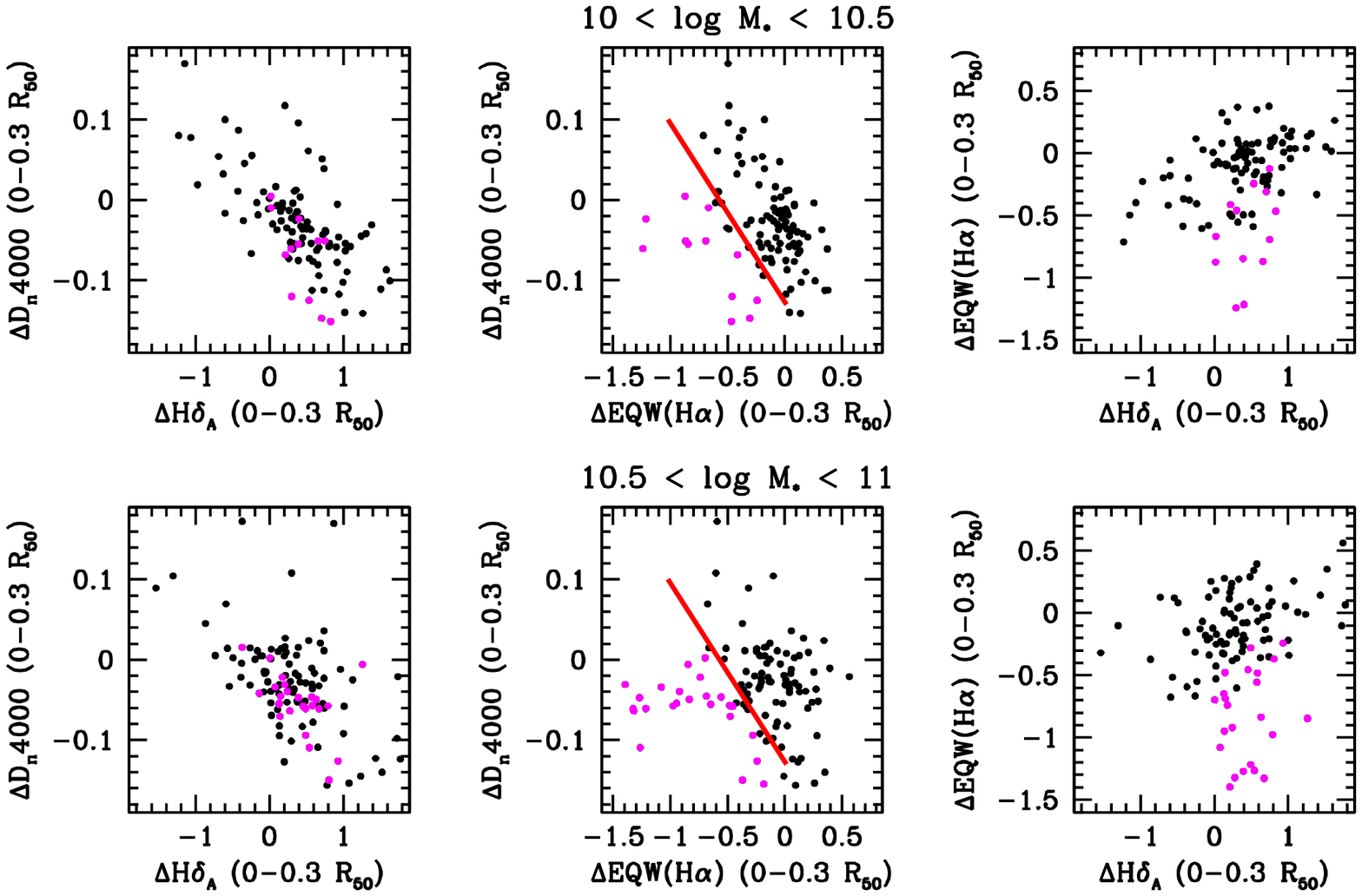}
\caption{ Correlations between the index differences for the central
0-0.3 R$_{50}$ radial bin. Results for galaxies in the stellar mass range
$10< \log M_*<10.5 M_{\odot}$ are shown in the upper panels and for 
$10.5< \log M_*<10.5 M_{\odot}$ in the lower panels.
The cut indicated by the
red line in the middle panels is used to define the outlier sample
with unusually steeply centrally rising EQW H$\alpha$ profiles.
\label{models}}
\end{figure*}

In the left panel of Figure 6,  individual spaxel measurements are plotted in the plane of log
EQW(H$\alpha$) versus D$_n$(4000) for  10 out of the 13  outlying galaxies shown as magenta
points in Figure 5.  \footnote {We do not show all objects to avoid overcrowing the
plot. We have selected the 10 objects with the most inner spaxel measurements 
that exhibit the largest dynamic range in EQW(H$\alpha$). These tend to be the most
nearby objects.}     The spaxel measurements are restricted  to those within
1.2 R/R$_{50}$ and where the signal to noise in both the H$\alpha$ and H$\beta$
line measurements are greater than 3, allowing for reasonably accurate
estimation of the Balmer decrement.  The plot includes spaxels for all outliers
in the stellar mass range $10 < \log M_* < 10.5$ Results for outliers in the
higher stellar mass bin are similar).  Central
(R$<$0.5 R$_{50}$) spaxel measurements from a subset of the outliers are plotted  as coloured
points. All points of the same colour come from the same galaxy.

\begin{figure*}
\includegraphics[width=150mm]{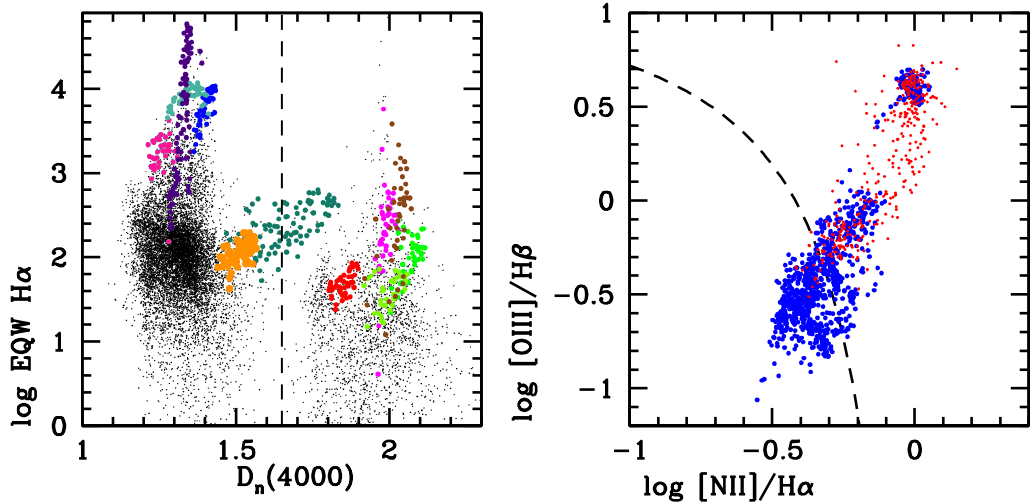}
\caption{ Left: Individual spaxel measurements for galaxies
in the stellar mass range $10<\log M_*<10.5M_{\odot}$ are plotted in the plane of log
EQW(H$\alpha$) versus D$_n$(4000) for the outlying galaxies shown as magenta
points in Figure 5.  The spaxel measurements are restricted to those within
1.2 R/R$_{50}$ and these are plotted as black points. 
Central
(R$<$0.5 R$_{50}$) spaxel measurements from a subset of the outliers are plotted  as coloured
points. All points of the same colour come from the same galaxy.
Right: Location of the central ((R$<$0.5 R$_{50}$) spaxels in the [OIII]/H$\beta$
versus [NII]/H$\alpha$  BPT diagram. Central spaxels with D$_n$(4000)$<1.6$ are
plotted in blue and  those with  D$_n$(4000)$>1.6$ are plotted in red.
\label{models}}
\end{figure*}

The central measurements form tight sequences  in the log EQW(H$\alpha$) versus
D$_n$(4000) plane, indicating that there is a strong and systematic offset in
the physical properties of the entire central region of the galaxy, not just in
a handful of isolated regions. It can also be seen that the spaxel measurements
separate into two main groupings at a D$_n$(4000) value of $\sim 1.6$, indicated
by a dashed line on the plot. In the right panel of Figure 6, the central 
spaxels are plotted  in the [OIII]/H$\beta$ versus [NII]/H$\alpha$ BPT diagram. Central spaxels
where D$_n$(4000)$<1.6$ are coloured in blue and those with  D$_n$(4000)$>1.6$
are coloured in red.     
As can be seen,  galaxies with central spaxel values of D$_n$(4000)$<1.6$
have low central ionization
parameters, indicating that the source of the excess H$\alpha$ emission must be
young stars rather than a central AGN. Although some of the spaxels lie just     
above the Kauffmann et al (2003) demarcation curve separating AGN from star-forming
galaxies, the fact that the H$\alpha$ equivalent widths are greater than 100
rule out ionization by evolved, post-AGB stars (Belfiore et al 2016).  The small cloud of blue points in the high
ionization region of the BPT diagram all come from a single galaxy, which is discussed in detail
in section 6.2  and is postulated to be a galaxy in transition between
star-forming and quiescence, because the H$\alpha$ line profiles exhibit
clear outflow signatures. The distribution of the red points in the BPT diagram shows
that a  much larger fraction of the galaxies with
D$_n$(4000)$>1.6$ have high central ionization parameters indicative
of AGN heating processes.  Because the interpretion
of these objects in terms of their stellar properties is less straightforward, 
these galaxies will not be considered further in  this paper and all further analysis
is restricted to the sub-sample of outliers with central spaxel values of D$_n$(4000)$<1.6$.

\section{Properties of the large H$\alpha$ EQW outlier galaxies}
This section characterizes the radial profiles of a sample of 15 galaxies with
outlying central stellar populations in the plane of  log EQW(H$\alpha$) versus
D$_n$(4000), where the average value of D$_n$(4000) for spaxels contained
within 0.3R$_{50}$ is less than 1.65. The galaxies are selected
from the full stellar mass range spanned by our parent sample. A ``control sample" of  15 galaxies with
similar central D$_n$(4000) values, stellar masses and redshifts, 
which are not classified as outliers in the
EQW(H$\alpha$)/D$_n$(4000) plane is extracted in order to characterize which aspects of the
radial trends are unique to the outlier population.

SDSS $g,r,i$ colour images of the outlier and control samples are shown in Figure
7. There are three clearly interacting galaxies among the outliers (top row,
columns 3 and 4, and bottom row, column 3). In addition, there are a number of
clearly lopsided  systems and galaxies with strong-armed spiral features 
suggestive of ongoing tidal
interactions.  In contrast, the control sample does not contain any interacting
systems and  consists of galaxies with more flocculent spiral structure.

\begin{figure}
\hspace{-7mm}
\includegraphics[width=95mm]{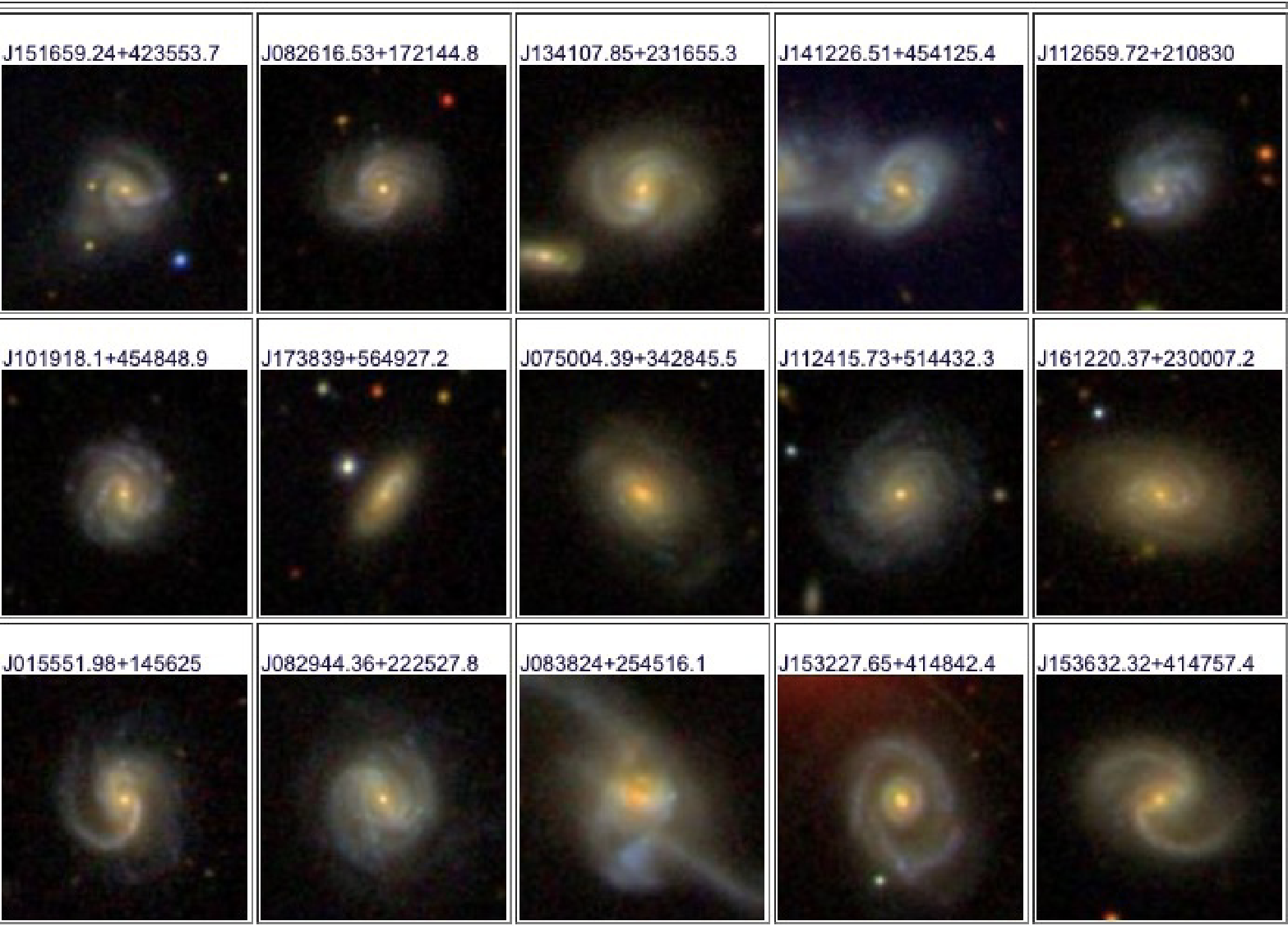}
\caption{ SDSS $g,r,i$ colour images of the outlier sample.} 
\hspace{-7mm}
\includegraphics[width=95mm]{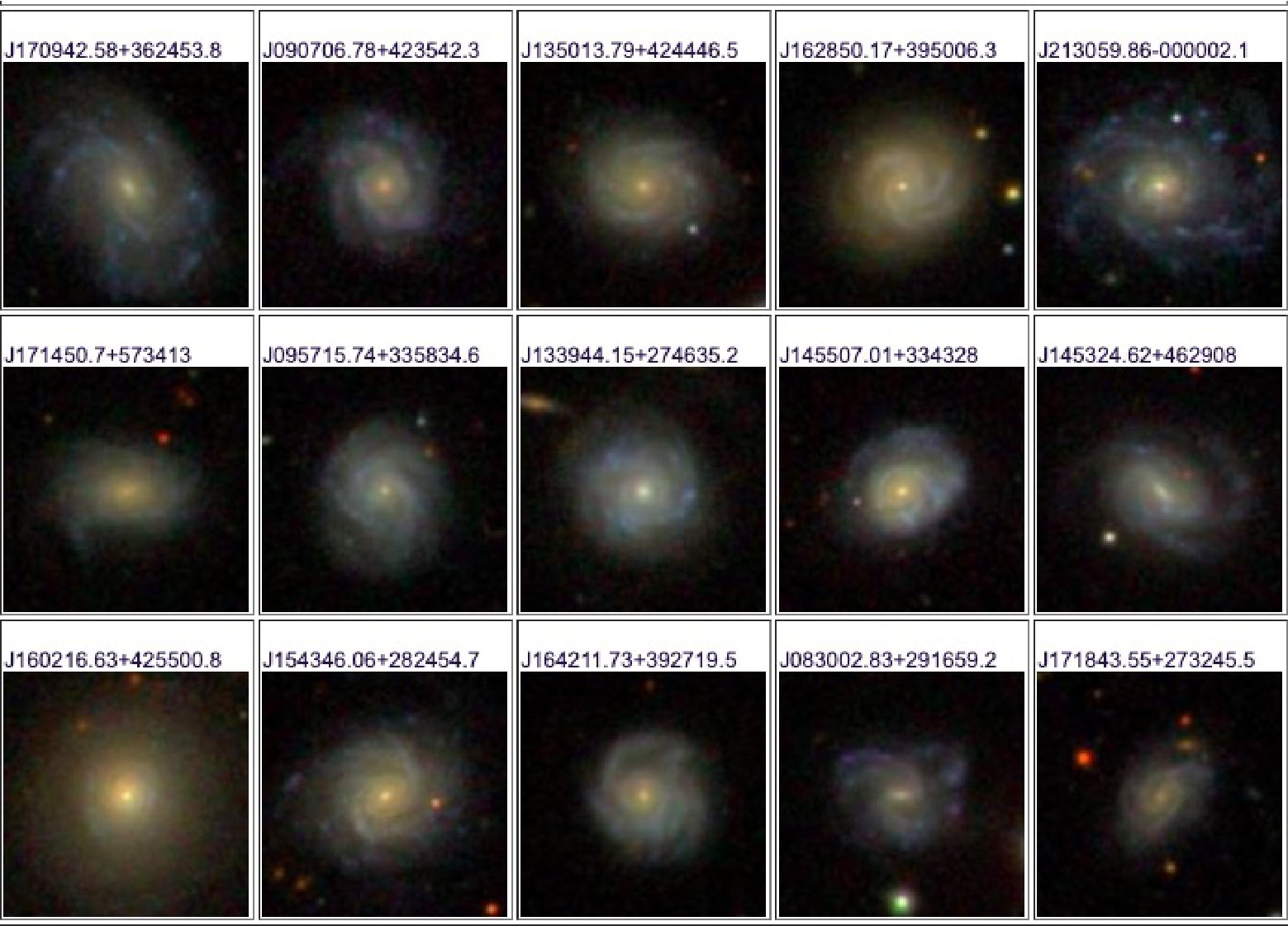}
\contcaption{ SDSS $g,r,i$ colour images of the control sample. 
\label{models}}
\end{figure}

In order to search for  signatures of the presence of young, massive stars, the
methodology decribed in Brinchmann et al (2008) has been implemented to identify
Wolf Rayet features in the spectra.  The main Wolf-Rayet features seen in the
optical spectra of galaxies are two broad emission features: the blue bump
around 4600-4680 \AA\ (rest-frame)  and the red bump around 5650-5800 \AA] 
(rest-frame).  Brinchmann et al
define an excess flux above the best-fit continuum in regions around the main
Wolf-Wayet features. The continuum fits provided by the DAP (see section 2) are
used to define the excess in the blue/red WR features as e$_{\rm blue/red}$=
F$_{\rm central,blue/red}$ -F$_{\rm continuum,blue/red}$ where F$_{\rm central}$ is the
summed flux around the central wavelength of each feature and F$_{\rm continuum}$ is
the summed flux from the continuum fit in two wavelength intervals around each
feature. The central wavelength positions  are given in Table
1 of Brinchmann et al. The central and continuum passbands both span a width of
100 \AA\ in total.

The blue Wolf Rayet feature includes the broad HeII  emission line at 4686 \AA\
that can sometimes be seen as a separate narrow feature. This is of potential
interest because binary evolution models (e.g. Eldridge \& Stanway 2009) predict
elevated HeII emission for a longer period of time than models
without binary stars.  Because the line is often
asymmetric, no attempt to fit a Gaussian is made, but the flux on either side of
4686 \AA\ is summed until the first measured flux point dips below the
continuum fit.

Finally, evidence of complex ionized gas kinematics is analyzed by searching for
the presence of multiple components in the  H$\alpha$ emission line profile.
The DAP includes the Gaussian model fit for each emission line after subtraction
of the best-fit continuum. The presence of outflowing gas is manifested as an
excess with respect to a single Gaussian fit on the blue side of the line
(see for example Forster-Schreiber
et al 2018). The reason we only look on the blue side is because a non-Gaussian component
on the red side of the line will be blended with [NII]$\lambda$6584.  
The excess is parametrized as F$_{\rm non-Gaussian}$, the fraction of
the total H$\alpha$ line flux shortwards of 6563 \AA\ that is in excess of the
single Gaussian fit provided by the DAP.

Figure 8 shows  montages of radial profiles for two examples of outlier
galaxies, while Figure 9 shows the same for two example control galaxies.
A minimum S/N cut of 3 is imposed on all emission line fluxes and stellar
continuum indices that are shown. Results are plotted out to a distance of 1.2
$R_{50}$ where the majority of spaxel measurements meet this criteria.
Magenta, red, green and blue points in each panel indicate
spaxels in the radial ranges 0-0.3 R$_{50}$, 0.3-0.6 R$_{50}$, 0.6-0.9 R$_{50}$, and
0.9-1.2 R$_{50}$, respectively.

\begin{figure}
\hspace{-7mm}
\includegraphics[width=91mm]{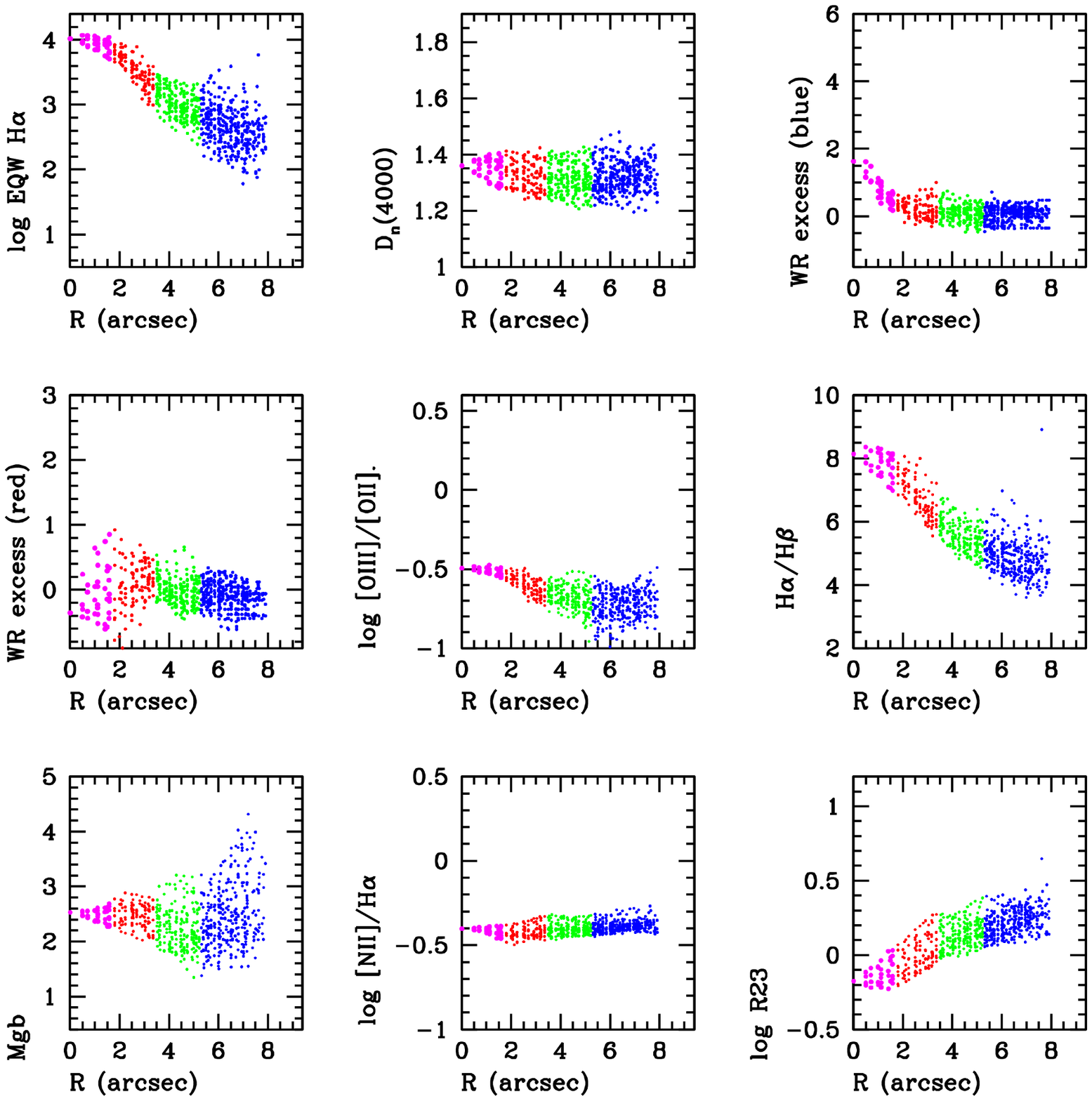}
\caption{Radial profiles for a variety of stellar and ionized gas disgnostics.
Magenta, red, green and blue points show spaxel located within  
0-0.3 R$_{50}$, 0.3-0.6 R$_{50}$, 0.6-0.9 R$_{50}$, and
0.9-1.2 R$_{50}$, respectively. These results are for the outlier galaxy in the
second row, third column in the first montage in Figure 7.}
\hspace{-7mm}
\includegraphics[width=91mm]{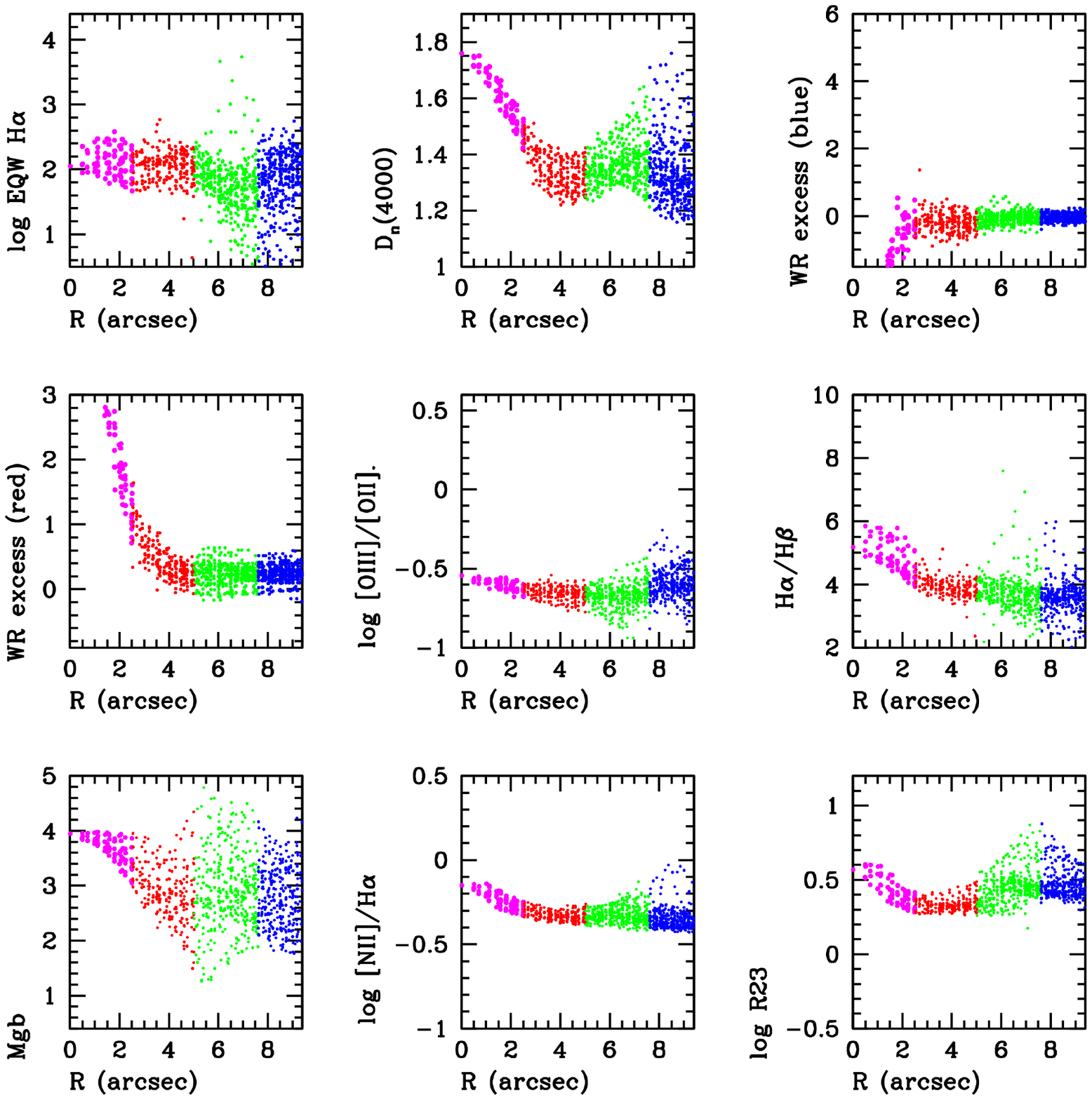}
\contcaption{ Same as the previous figure, except for the outlier galaxy shown in  
third row, fourth column in the first montage in Figure 7.
\label{models}}
\end{figure}

\begin{figure}
\hspace{-7mm}
\includegraphics[width=91mm]{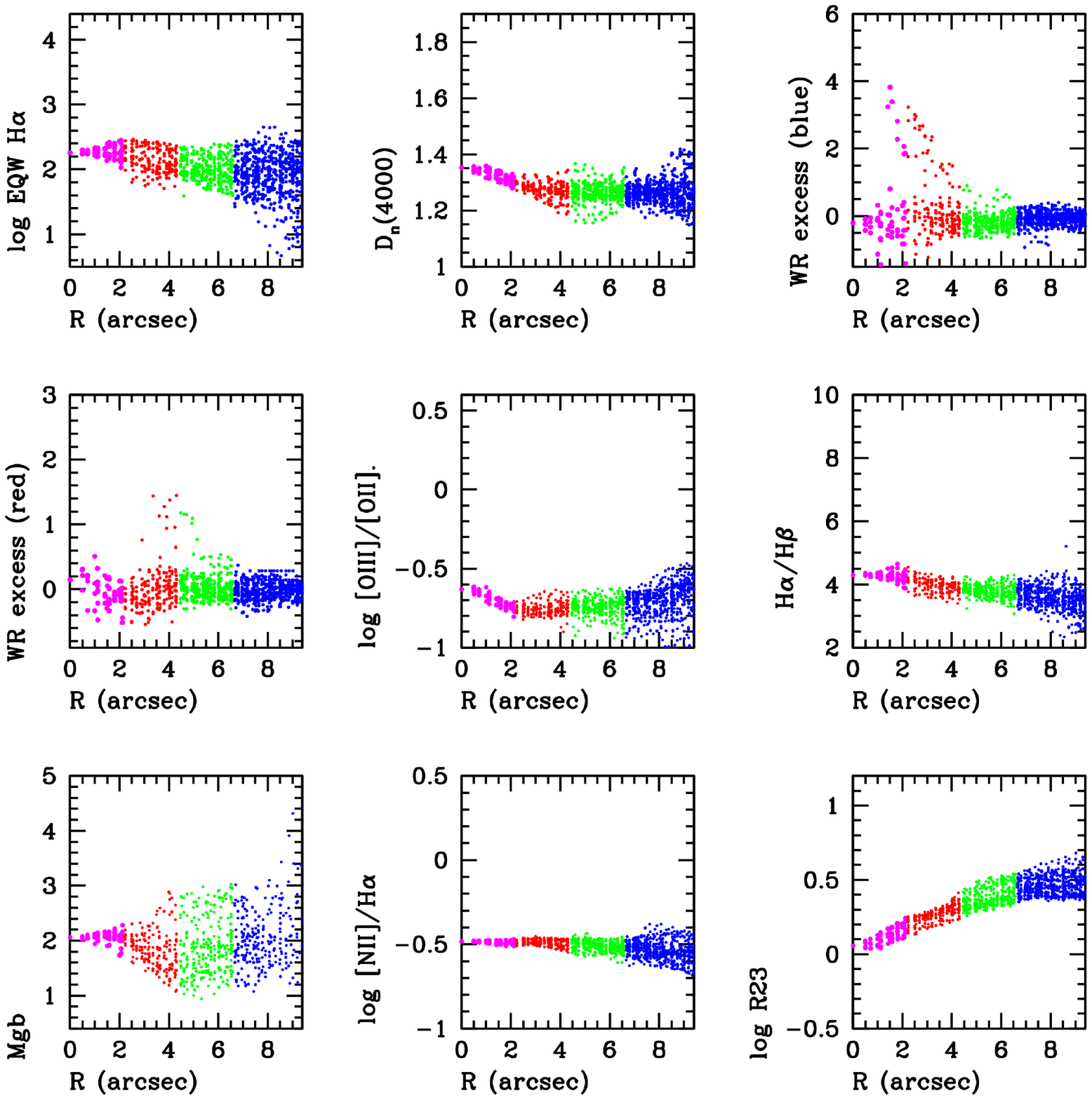}
\caption{Radial profiles for a variety of stellar and ionized gas disgnostics.
Magenta, red, green and blue points show spaxel located within  
0-0.3 R$_{50}$, 0.3-0.6 R$_{50}$, 0.6-0.9 R$_{50}$, and
0.9-1.2 R$_{50}$, respectively. These results are for the control galaxy in the
top row, first column in the second montage in Figure 7.}
\hspace{-7mm}
\includegraphics[width=91mm]{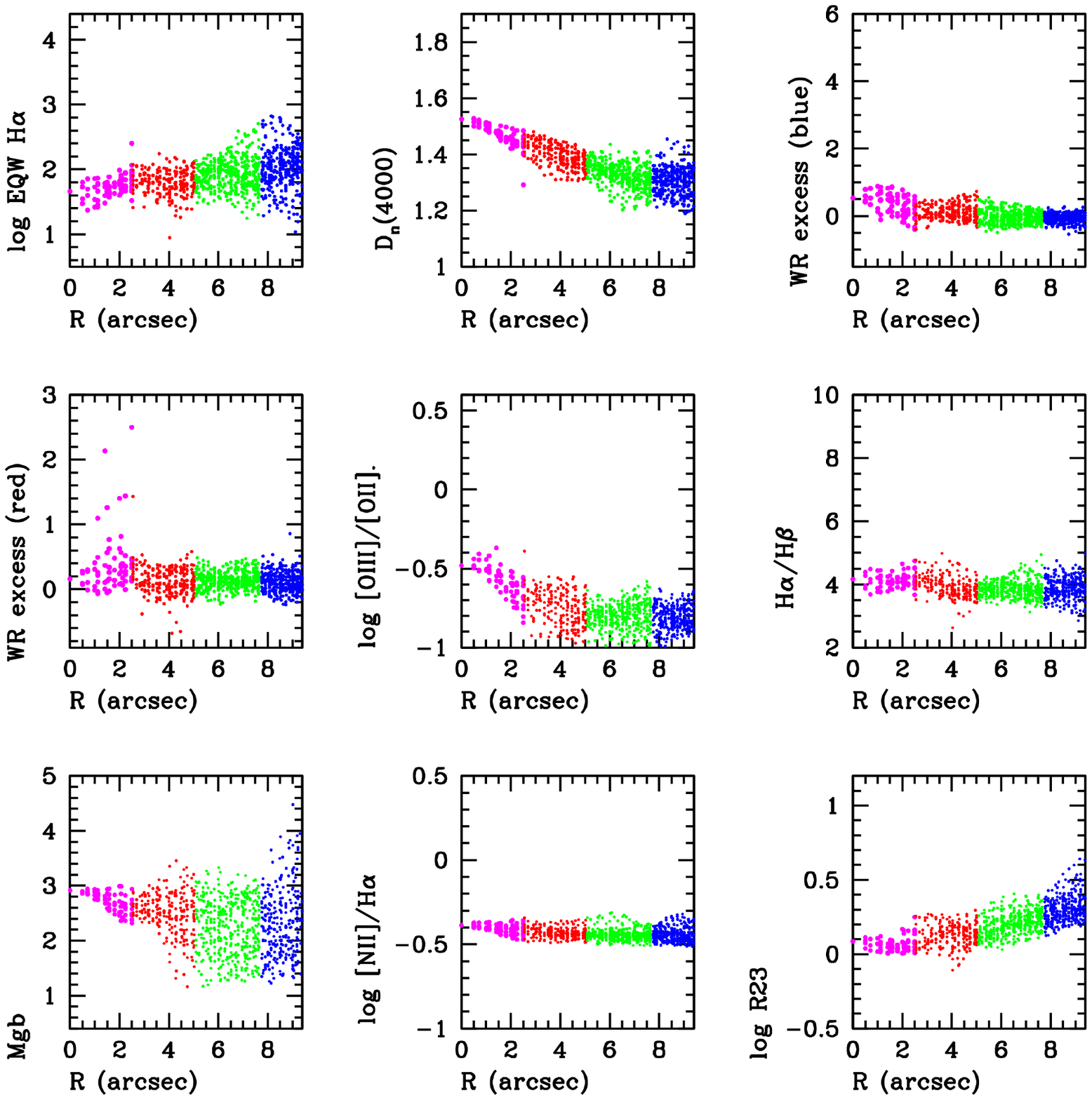}
\contcaption{ Same as the previous figure, except for the control galaxy shown in  
top row, third column in the second  montage in Figure 7.
\label{models}}
\end{figure}

Significant variations in profiles are found between different galaxies from the
two samples. The second  galaxy shown in Figure 8 has a flat H$\alpha$ equivalent width
profile all the way from the central region to 1.2 R$_{50}$, but D$_n$(4000)
drops steeply out to 0.5$R_{50}$ and then flattens. No blue Wolf Rayet excess is
detected at any radius, but the red Wolf Rayet excess increases towards the
central regions of the galaxy at $R<$ 0.5$R_{50}$. The galaxy  has a rather flat
ionization parameter ([OIII]/[OII]) profile and a Balmer decrement profile
than rises gently in the inner region to a value of $\sim$ 6.
The gas-phase metallicity, as measured by the parameter R23=
([OII]$\lambda$3727+[OIII]$\lambda$5007)/H$\beta$, and the
stellar absorption line index Mgb, which is
most sensitive to the alpha element Magnesium, both have  flat
gradients across the galaxy.  In contrast, the first galaxy shown in Figure
8  has a H$\alpha$ equivalent width profile that rises steeply
towards the central regions of the galaxy, and a D$_n$(4000) profile that is
flat. A small central blue Wolf Rayet excess is detected, but no central red
excess is found. The galaxy also has a very steeply centrally
rising  Balmer decrement, but a gas-phase metallicity gradient that
decreases towards the central regions. 

The two control sample galaxies shown in Figure 9 show a similar degree of
variability, especially in stellar and gas-phase metallicity indicators.
It is thus clearly important to systematize the comparison between the outlier
galaxies and the control sample.
Values in the central spaxels  are recorded for quantities that are independent of
the selection procedure (e.g. [OIII]/[OII]).   
Emission line and stellar absorption index
profiles are classified according to whether the quantity under study rises or
falls towards the central regions of the galaxy, or is flat. Because the Wolf
Rayet excess parameters, the HeII line equivalent widths and the non-Gaussian
H$\alpha$ line shape measures are based on weak features in the spectrum and are
more sensitive to systematic problems such as offsets in the stellar continuum
fits, the galaxy is classified according to whether it exhibits a clear
excess in these measurements in the  central region compared to the outskirts.

The main results are summarized as
follows:
\begin{enumerate}
\item {\bf log H$\alpha$ equivalent width:}\\ {\em Outliers:} 13 rising, 2 flat\\  {\em Controls:}
2 rising, 6 flat, 7 falling
\item {\bf D$_n$(4000):}\\{\em Outliers:} 12 rising, 2 flat, 1 falling\\  {\em Controls:}
9 rising, 6 flat
\item {\bf log H$\alpha$/H$\beta$:}\\ {\em Outliers:} 13 rising, 2 flat\\ {\em Controls:}
3 rising, 12 flat
\item {\bf log [OIII]/[OII]:}\\ {\em Outliers:} 13 rising, 2 flat, median
central value: -0.5\\  {\em Controls:} 7 rising, 7 flat, 1 falling, median
central value: -0.6
\item {\bf Blue WR central excess:}\\ {\em Outliers:} 6 yes, 9 no\\  {\em Controls:}
3 yes, 12 no
\item {\bf Red WR central excess:}\\{\em Outliers:} 12 yes, 3 no\\   {\em Controls:}
4 yes, 11 no
\item {\bf  HeII central excess:}\\ {\em Outliers:} 3 yes, 12 no\\  {\em Controls:}
3 yes, 12 no
\item {\bf non-Gaussian H$\alpha$ central excess:}\\ {\em Outliers:} 1 yes, 14 no\\  {\em Controls:}
0 yes
\item {\bf Mgb Lick index:}\\{\em Outliers:} 11 rising, 4 flat, 
median central value: 2.5\\  {\em Controls:} 6 rising, 5 flat, 4 falling,
median central value: 2.0
\item {\bf R23 index:}\\{\em Outliers:} 4 rising, 4 flat, 7 falling 
median central value: 0.25\\  {\em Controls:} 2 rising, 3 flat, 10 falling,
median central value: 0.0
\item {\bf log [NII]/H$\alpha$:}\\ {\em Outliers:} 7 rising, 8 flat, median
central value: -0.3\\  {\em Controls:} 3 rising, 12 flat, median
central value: -0.5
\item {\bf VLA FIRST radio detection:}\\ {\em Outliers:} 8 yes, 7 no\\ 
{\em Controls:} no detections
\end {enumerate}

Findings (i), (ii)  and (iii)  are direct consequences of the selection procedure used to
define the outlier sample. Taken together, these three results show that the
outlier sample consists of galaxies with more centrally peaked H$\alpha$
emission and that these regions are strongly obscured by dust. Finding (iv),
showing that the central ionization parameter is only marginally higher for the
outliers compared to the controls, indicates that young stars are the most
probable central ionizing sources in both samples. The fact that the ionization
parameter is more often centrally rising in the outliers, indicates a possible
systematic change in the nature of the ionizing stars with radius in these
galaxies.

The suggestion that the outliers host a systematically different population of
young stars in their centres is further strengthened by findings (v) and (vi), which
show that a central Wolf Rayet excess is more frequently found in the outliers.
The fact that the red Wolf Rayet excess is more common than the blue Wolf Rayet
excess is particularly intriguing, because it indicates that these central
starbursts are very different to those in UV-selected Wolf Rayet galaxies; this
will be discussed in more detail in the next section.

Findings (vii) and (viii) are concerned with very weak spectral features that are at the
limit of detectability. The results indicate that the [HeII]$\lambda$4686 line
is not the dominant contributor to the  blue Wolf Rayet excess in all the
outliers.  The lack of clear
non-Gaussian signatures in the H$\alpha$ line shows that if there are  accreting
black holes hidden in the central regions of these galaxies, they are not usually driving noticeable
outflows of  ionized gas. The case of the outlier with a clear H$\alpha$
outflow signatures will be discussed in detail in the next section.

Findings (ix), (x) and (xi)  show that the outliers more often have centrally rising gas-phase
and stellar metallicities compared to the control galaxies, and that their central
metallicities  are also systematically higher. This accords with
finding (iii) that the outliers have higher central dust content (dust attenuation 
and metallicity are usually quite strongly correlated in star-forming systems).

Finally, finding (xii)  that  53\%  of the outliers have detectable 20 cm radio
emission compared to 0\% of the control galaxies deserves some comment. The
radio fluxes of the detected galaxies range from 1 mJy (the detection limit of
FIRST) to 90 mJy, with a median value of 2.2 mJy. Even the brightest sources are
unresolved at the 5 arcsec resolution of the FIRST observations. This again
supports the conclusion that the central sources in the outliers are likely
stellar rather than AGN.

\section{Discussion}
In this section, a number of broader issues and unanswered questions arising
from this work are discussed.

\subsection{Inferences from stellar population synthesis models} The analysis so
far has been been empirical in approach. Starting from a sample of $\sim$700
galaxies, the locus occupied by disk stellar populations in the plane of
H$\delta_A$ versus D$_n$(4000) and H$\alpha$ equivalent width versus D$_n$(4000)
is defined. Galaxies with central stellar populations that deviate strongly are
identified. Once the outlier sample has been selected, radial trends for a  variety
of stellar and ionized gas diagnostics are studied in order to show that
signatures from very young massive stars become progressively stronger
towards the very centers of these galaxies. Because H$\alpha$
is sensitive to the presence of only the most massive
stars, these results suggest that stars may be  
forming in these systems with an initial mass function (IMF) that is
significantly different to that in galactic disks.

Stellar population synthesis modelling is, in principle, an important way to
test whether these inferences accord with theoretical expectations and to
quantitatively constrain parameter changes in the IMF. In practice, models of
massive star evolution have many uncertain aspects,  such as the
effect of rotation, treatment of binary stars and their evolution, mass loss,
and so on (see for sample Smith (2014) for a review). 
I will not delve very deeply into these matters in this paper, but
will show some illustrative comparisons using the publically available Starburst
99 code (Leitherer et 1999; 2014) to predict the H$\alpha$ equivalent width of
galaxies and models from Bruzual \& Charlot (2003) to predict the stellar
continuum indices D$_n$(4000) and H$\delta_A$.

In  Starburst 99,
it is necessary to combine a variety of inputs from very heterogeneous sources.
The 4000 \AA\ break and H$\delta_A$ predictions are generated using the high
resolution stellar library in Martins et al (2005) consisting of synthetic
stellar spectra. The predicted relation between D$_n$(4000) and H$\delta_A$
differs significantly from other models that use empirical libraries
(e.g. Bruzual \& Charlot 2003)  and is also
significantly offset from the data shown in Figures 1 and 2. 
Unfortunately, the models that generate better
fits to the stellar continuum features do not include detailed predictions for
nebular features in the spectra. 
In this work, predictions for H$\alpha$ luminosity evolution for a simple stellar population
(SSP) as a function of time are taken from Starburst99 and are are matched to
the evolution of the stellar continuum fluxes from the Bruzual \& Charlot (2003)
models at the same output times.
The top two panels of Figure 10 show locii occupied by galaxies in the
H$\delta_A$ versus D$_n$(4000) (left) and EQW  H$\alpha$ versus D$_n$(4000)
planes (right) occupied by  galaxies that have had 
continuous star formation histories.  All
model galaxies begin forming stars 9 Gyr in the past and their star formation
rate is parametrized as $SFR(t)= C e^{(\gamma t)}$, where $\gamma$ varies over the
range -2.0 to 0.0, i.e. from steeply declining to constant
star formation rate as a function of time. Black, green and blue curves show results for solar,
0.5 solar and 0.25 solar metallicity models. A standard Kroupa (2001) initial
mass function has been adopted. This IMF has the standard
Salpeter slope $\alpha = 2.3$ above half a solar mass, 
but introduced $\alpha = 1.3$  between 0.08-0.5 $M_{\odot}$
and  $\alpha= 0.3$ below 0.08 $M_{\odot}$. This IMF is widely adopted when fitting
normal galaxy stellar populations because it provides simultaneous good fits
to the observed mass-to-light ratios and colours of
nearby spiral galaxies (e.g. McGaugh \& Schombert 2014).

\begin{figure}
\includegraphics[width=94mm]{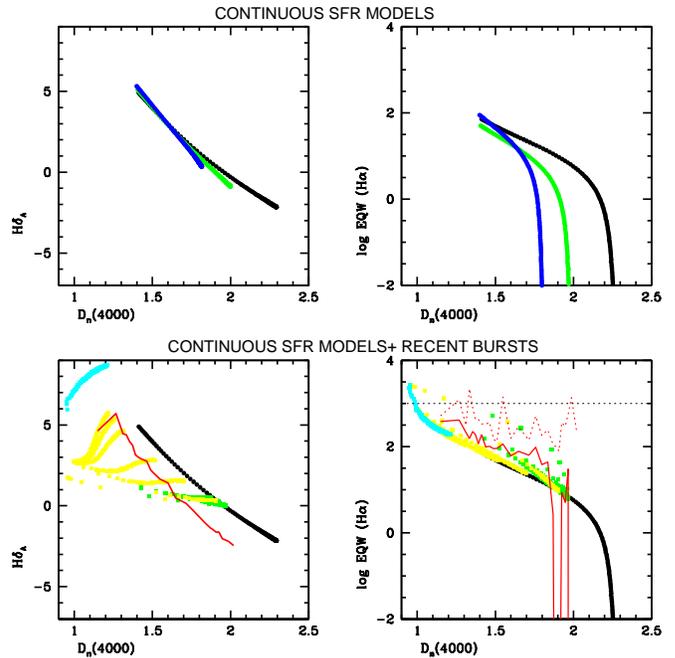}
\caption{ Top panels: Locii occupied by galaxies in the
H$\delta_A$ versus D$_n$(4000) (left) and EQW  H$\alpha$ versus D$_n$(4000)
planes that have had continuous star formation histories.        
Black,green and blue curves show results for solar,
0.5 solar and 0.25 solar metallicity models.
Bottom panels: Model with recent bursts superposed upon the grid of models
with continuous star formation histories and solar metallicity (black curve).
The yellow points
are for strong bursts that contribute between 0.1 and 50\% of the total mass in
stars. The green points are for weak bursts that contribute between 0.005 and 0.1\%
of the mass.
The cyan points show
the location of  ``pure" starburst model with no underlying older stellar population. 
The solid and dotted red curves show the median and 90th percentile relations
shown in the two left-hand panels of Figure 3 for the inner
stellar populations of low mass galaxies with $10 < \log M_* < 10.5$.
\label{models}}
\end{figure}

The bottom two panels show how recent bursts superposed upon the grid of models
with continuous star formation histories and solar metallicity influence  the
predicted location in the two planes.  For reference, the black  curve shows the same solar
metallicity continuous models plotted in the upper panels. The cyan points show
the location of  ``pure" starburst models. The pure starbursts begin forming
their stars between 5$\times10^8$ years and $10^6$ years ago and form stars at a
constant rate.  The yellow and green points show the location of continuous
models with superposed pure starbursts of differing amplitude. The yellow points
are for strong bursts that contribute between 0.1  and 50\% of the total mass in
stars. The green points are for weak bursts that contribute between 0.005 and 0.1\%
of the total stellar mass.  Superposing an ongoing starburst onto an underlying
continuous model  moves galaxies to lower values of D$_n$(4000). Interestingly, the
starburst moves the galaxy below the
continuous model locus in the H$\delta_A$ versus D$_n$(4000) plane, but moves
the galaxy {\em along this locus} in the EQW H$\alpha$ versus D$_n$(4000) plane.
The small smattering of points that are dispaced well above the locus are for
extremely young bursts that are less than a few million years old superposed on
relatively old stellar populations.

Note that even with very young bursts, it is not possible to reach H$\alpha$
equivalent width values close to a thousand at intermediate values of
D$_n$(4000), as observed in the central regions of some of the outlier galaxies.
For reference, we have plotted the  median and 90th percentile relations
shown in the two left-hand panels of Figure 3 for the inner
stellar populations of low mass galaxies with $10 < \log M_* < 10.5$ as red solid and
dotted lines. As can be seen, the 90th percentile in the distribution of
EQW (H$\alpha$) at fixed D$_n$(4000) is not reachable even for extremely young bursts,
particularly at the larger D$_n$(4000) values.  

I now turn to a very simple first exploration of the effect of changing the
upper mass end of the IMF. In the Galactic Center, the upper-end of the IMF
slope has been constrained using the K-band luminosity function of luminous
early-type stars located in two disk-like structures in the inner 0.5 pc of the
Galactic bulge. Paumard et al (2006) obtain a best fit mass function with
a considerably flatter slope than the standard Salpeter one ($dN/dm=m^{-0.85}$
compared to $dN/dm=m^{-2.35}$). Because of the very young inferred ages of the
stars, it was concluded that most of the stars must have formed in place from
gas in the observed disks, rather than having migrated  inwards from birth clouds at
larger radius. In more recent work, Lu et al (2013) estimate a slope
$dN/dm=m^{-\alpha}$, with $\alpha= 1.5-1.9$ for young stars with masses above 10
$M_{\odot}$, considerably flatter than Salpeter but not as extreme as that
estimated in earlier work.

Motivated by these results, Figure 11 explores the effect of flattening the IMF
slope away from the Salpeter value above some characteristic stellar mass.  The
purpose here is to examine the outer boundaries of the possible  variations in
H$\alpha$ equivalent width under the most extreme IMF changes. Results from two
models are compared with the fiducial case of solar metallicity continuous star
formation models with Salpeter IMF (red curves) : \begin {enumerate} \item  a
model where $\alpha=0.5$ above 7 $M_{\odot}$ (blue curves) \item  a model where
$\alpha=0.8$ above 10 $M_{\odot}$ (green curves) \end {enumerate} As can be
seen,  these models result in H$\alpha$ equivalent widths that are a factor 4-7
higher at a fixed value of D$_n$(4000). Wolf Rayet emission features are 
predicted by the Starburst99 code, but it is difficult to combine these
with the stellar continuum predictions in a robust way, so we will defer 
detailed examination of red and blue bump predictions to future work.

\begin{figure}
\includegraphics[width=90mm]{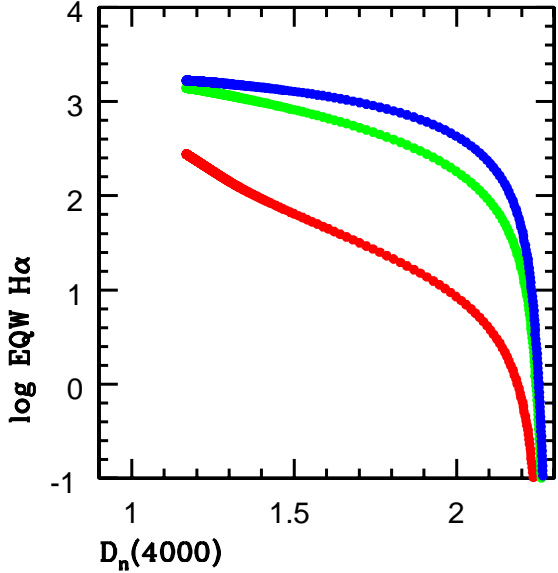}
\caption{ Locii occupied by galaxies in the
EQW  H$\alpha$ versus D$_n$(4000)
plane that have had continuous star formation histories.                             
The red curve shows results for the fuducial IMF, the green and blue
curves show results for the two IMF changes described in the text
(green is for case (i) and blue id for case (ii)).  
\label{models}}
\end{figure}

Another issue that should be addressed in future work is the dust correction.  Almost
all the galaxies in the outlier sample have Balmer decrement profiles that rise
towards the central region of the galaxy.  This means that the corrections that
have to be made in order to estimate the true production rate of ionizing
photons become progressively larger at smaller radii, which will complicate
quantitative interpretation of the results.  One might also ask whether some
fraction of the massive star population could simply be so deeply embedded in
dust so as to be completely invisible at optical wavelengths.

One important diagnostic of the clumpy nature of the dust in the
central regions of the outlier galaxies in our sample, is the scatter in the
Balmer decrement values measured for different spaxel spectra at the same
radius.  The scatter in H$\alpha$/H$\beta$ at fixed R/R$_{50}$ is small in all
the examples shown in Figures 8 and 9. Only one galaxy in our sample is
substantially different -- this is the recent major merger remnant pictured in
the bottom row, third column of the outlier galaxy image
montage shown in   Figure 8. The H$\alpha$ equivalent width and
Balmer decrement profiles for this galaxy are shown in Figure 12. The scatter in
both is very much larger in this galaxy. The Balmer decrement values appear to
lie along two tracks, a high and a low attenuation one, perhaps
indicating incomplete mixing of the interstellar medium components of the two
progenitor galaxies that merged to form this object.

\begin{figure}
\includegraphics[width=100mm]{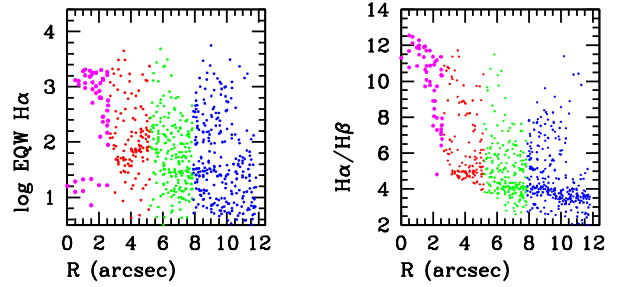}
\caption{The H$\alpha$ equivalent width and
Balmer decrement profiles for the recent major merger remnant pictured in
the bottom row, third column of the outlier galaxy image
montage shown in  Figure 8.
\label{models}}
\end{figure}

\begin{figure}
\includegraphics[width=76mm]{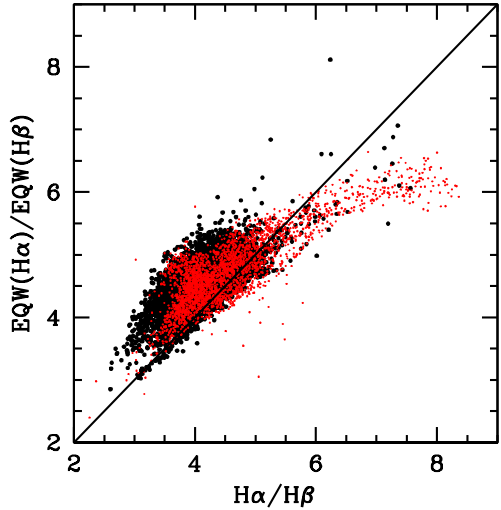}
\hspace{-9mm}
\caption{The ratio EQW(H$\alpha$)/EQW(H$\beta$), which includes the measured stellar
continuum at the wavelengths of the two Balmer lines, is plotted as a function
of the standard Balmer decrements H$\alpha$/H$\beta$ for spaxels with
R/R$_{50}$$<0.5$ for control galaxies (black) and for outliers (red).
\label{models}}
\end{figure}

Even if the dust distribution in the centers of most of the outlier galaxies is
homogeneous, in order to compare the observational measurements with population
synthesis model predictions, both the Balmer line emission and the stellar
continuum fluxes have to be properly dust-corrected. Figure 13 shows how such
corrections vary systematically between the outliers and the control sample. The
ratio EQW(H$\alpha$)/EQW(H$\beta$), which includes the measured stellar
continuum at the wavelengths of the two Balmer lines, is plotted as a function
of the standard Balmer decrements H$\alpha$/H$\beta$ for spaxels with
R/R$_{50}$$<0.5$ for control galaxies (black) and for outliers (red). The two
samples largely overlap each other for Balmer decrement values up to $\sim 6$.
The outlier sample shows a tail of Balmer decrement values extending out to
values in excess of 8, where the ratio EQW(H$\alpha$)/EQW(H$\beta$) appears to
saturate at a fixed value.  These high dust attenuation spaxels originate from the
very central regions of the outliers. Once again, more detailed modelling via
full spectrum-fitting techniques (e.g. Wilkinson et al 2017) is required for more
detailed interpretation and will be the subject of future work.

\subsection{Growth of black holes and the AGN connection} Dense inner galactic
bulges with an excess population of young massive Wolf Rayet stars are likely to
be very interesting sites for studying black hole growth and accretion
processes. The most massive of these stars will form stellar black holes
at the end of their evolution and will also expel substantial material
in winds that may later cool and accrete onto the central black hole
if it is present. Large populations of stellar mass black holes may
also merge in dense environments to form a new black holes. 
 Within the context of the IFU survey at hand, further progress in
this area could be made by searching for and studying {\em transition objects},
i.e. galaxies with extreme central starbursts that will later become
AGN.

Recall that the outlier sample is selected in the plane of EQW H$\alpha$ versus
D$_n$(4000) and is restricted to galaxies with central values of D$_n$(4000)
less than 1.6, i.e. to bulges with young to middle-aged stellar populations.
There is one galaxy among the sample of 15 that has significantly higher ionization
parameter than the other outlier galaxies and may plausibly be considered as an
example of a transition object.  This is the interacting spiral shown  in the
top row, 3rd column of Figure 7; it's traditional name is IC 0910, classified as a
LINER in the catalogue of Veron-Cetty \& Veron (1996).

Figure 14 shows a compilation of radial profiles for this object. 
Note that this galaxy is among those with the strongest Wolf Rayet signatures.
It is unusual in that it shows strong blue bump and red bump excesses,
significant central [HeII]$\lambda$4686 emission and clear non-Gaussian
H$\alpha$ profiles in the centre of the galaxy.  Most of the outliers only
exhibit a subset of these signatures.  It is also the outlier where the
profiles of stellar and ionized gas metallicity and dust rise most strongly near
the centre of the galaxy and the central BPT line ratios are consistent with
AGN rather stellar ionization.

\begin{figure}
\hspace{-7mm}
\includegraphics[width=91mm]{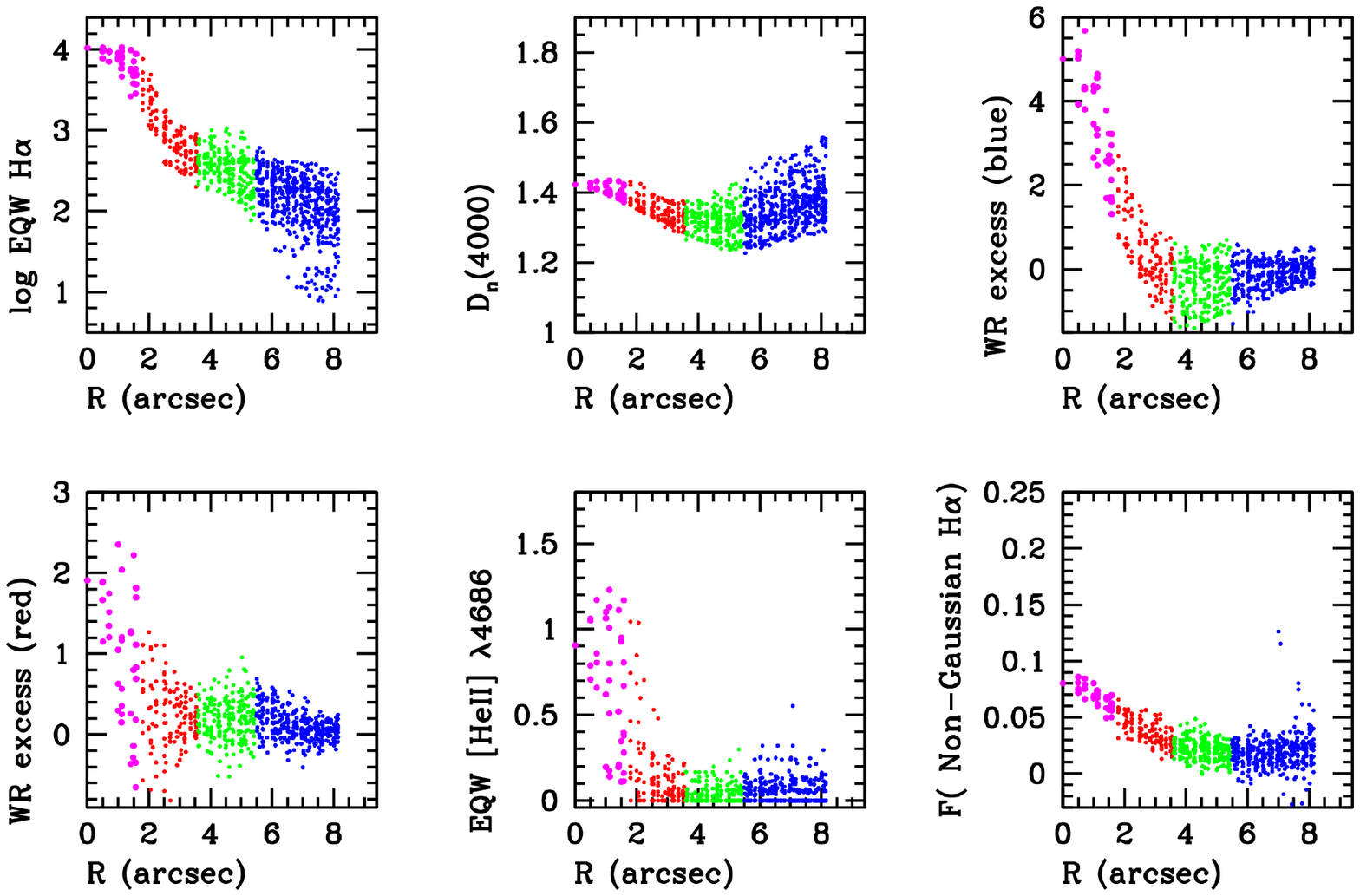}
\caption{Radial profiles for a variety of stellar and ionized gas disgnostics.
Magenta, red, green and blue points show spaxel located within  
0-0.3 R$_{50}$, 0-0.3 R$_{50}$, 0-0.3 R$_{50}$, and
0-0.3 R$_{50}$, respectively. These results are for the outlier galaxy in the
tope  row, third column in the first montage in Figure 7.}
\hspace{-7mm}
\includegraphics[width=91mm]{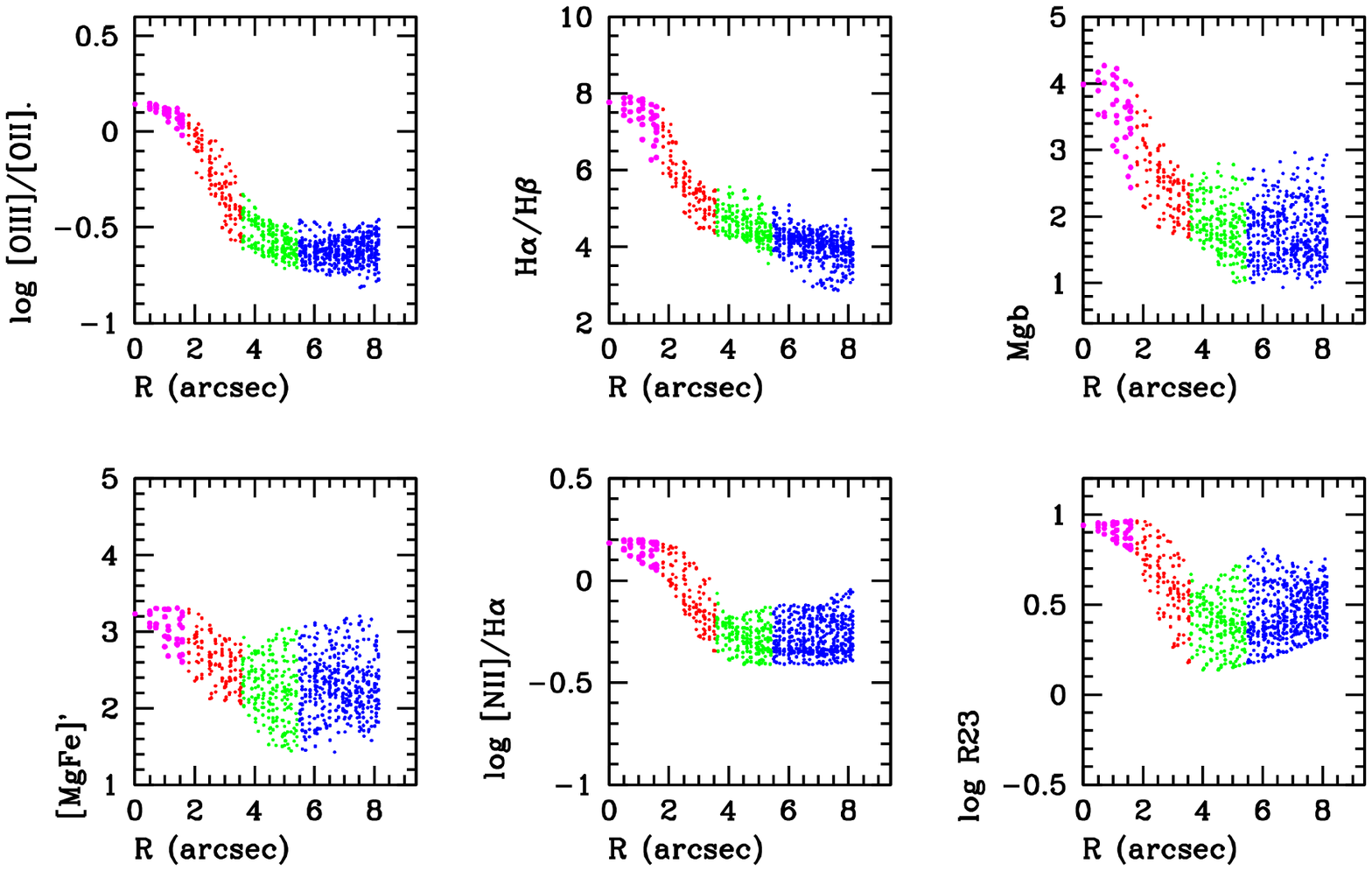}
\contcaption{ More radial profiles.     
\label{models}}
\end{figure}

One question is whether the observed non-Gaussian signature in the H$\alpha$
emission line is indicative of the presence of an outflow. Another
question concerns the extent to
which the outflow is influencing the gas in the galaxy. A first attempt to
answer this question is made in Figure 15 where  the ratio of the
velocity dispersion of the H$\alpha$ emitting gas to the stellar velocity 
dispersion is plotted as a function of log EQW H$\alpha$ for individual spaxel
spectra. The black points show measurements from spaxels located within 0.3
R$_{50}$ for the control galaxies and red points are for the outlier galaxies.
Spaxels from the galaxy highlighted in this section are coloured as magenta
points.  From this, it can be seen that the galaxy selected as a
possible transition object is unique. The ratio $\sigma$(gas)/$\sigma$(stars)
lies between 0.1 and 1 for almost all the measurements in the central regions of
both the outlier and the control sample. For the transition object,
$\sigma$(gas)/$\sigma$(stars) rises to values of 1.5-1.7 at the very center of
the galaxy. This indicates that feedback from a centrally accreting black hole
or from winds generated by massive stars is stirring up the gas very
significantly in this particular system.  It would obviously be very useful to
increase  sample sizes in future work and study whether it is possible to map
out a clear evolutionary sequence in central stellar population properties, as
well as in gas and stellar kinematics. In recent work, D'Agostino et al (2019)
suggest that resolved observations of the  velocity dispersion of gas in AGN can be used as a diagnostic
of where shock-heating processes are important.

\begin{figure}
\includegraphics[width=76mm]{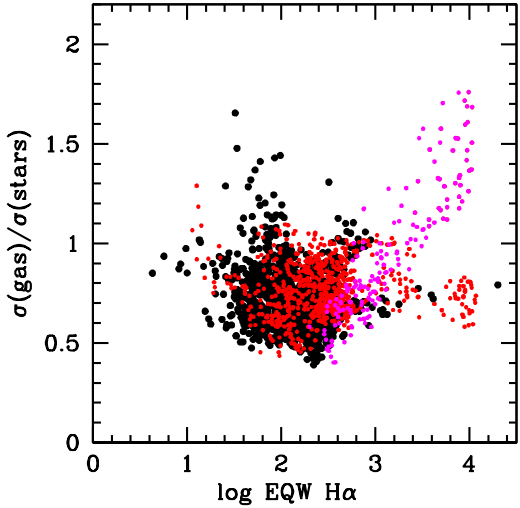}
\hspace{-9mm}
\caption{The ratio of the
velocity dispersion of the H$\alpha$ emitting gas to the stellar velocity
dispersion is plotted as a function of log EQW H$\alpha$ for individual spaxel
spectra. The black points show measurements from spaxels located within 0.3
R$_{50}$ for the control galaxies and red points are for the outlier galaxies.
Spaxels form the galaxy highlighted in this section are coloured as magenta
points.
\label{models}}
\end{figure}

\subsubsection{A more detailed examination of Wolf Rayet features} W-R galaxies are
usually defined as those galaxies in whose integrated spectra a broad emission
feature at He II $\lambda$4686 attributable to Wolf-Rayet stars has been detected
(Conti, 1991). W-R galaxies identified this way have been found to span the full
range of stellar masses and morphologies as the ordinary galaxy population, but
are exclusively found in galaxies with the bluest colors and highest star formation rates
for their mass (Brinchman et al 2008; Liang et al 2020). A blue bump
selection method has been demonstrated to produce samples where detection of the red bump is
relatively rare (Schaerer \& Vacca  1998; Brinchmann et al 2008).

\begin{figure*}
\includegraphics[width=141mm]{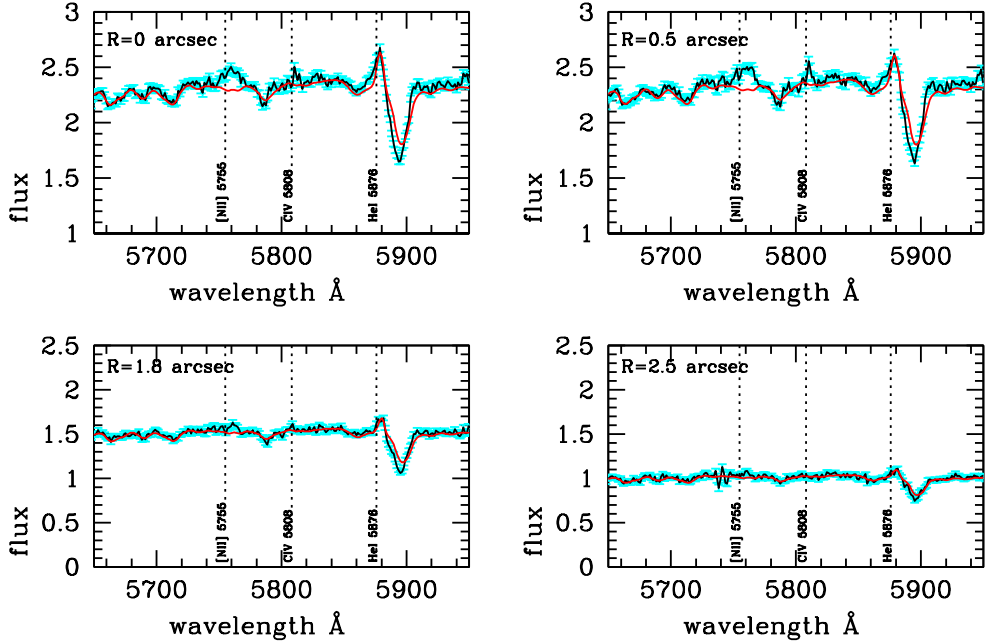}
\caption{A series of 4 spaxel spectra over the wavelength range spanning
the red bump Wolf Rayet feature for the same galaxy as in the previous figure.   
The spectra are arranged in order of increasing distance from the centre of the
galaxy. The positions of the the main emission line features are indicated in each
panel. The C IV $\lambda$5808 line is clearly detectable in the first two 
central spaxel spectra
and shows a broad component. 
\label{models}}
\end{figure*}

It is thus intriguing that the frequency with which the red bump is detected in
the outlier sample is considerably higher than for the blue bump. A detailed
physical interpretation of this finding lies beyond the scope of this paper.

One question is whether the red bump can be clearly resolved into individual
emission lines, which might assist in further interpretation of this feature.
Wolf-Rayets stars are divided into 3 classes based on their spectra, 
the WN stars (nitrogen dominant, some carbon), WC stars (carbon dominant, no nitrogen), 
and the rare WO stars with C/O$<1$.
The  broad C IV $\lambda$5808
emission line dominates the red bump and its strength
relative to the blue bump should be indicative of the  WC-to-WN star ratio. 
Wolf-Rayet star features could, in principle, be 
very IMF sensitive.  More detailed research  is required, however,
to properly constrain the progenitor masses of different Wolf-Rayet classes. 
Whether or not the  WC-to-WN star ratio can be regarded as a
constraint on the IMF or a disgnostic of the formation rate
of stellar mass black holes remains a  
topic of controversy (Sander et al 2019).

Figure 16 presents a series of 4 spaxel spectra over the wavelength range spanning
the red bump Wolf Rayet feature for the galaxy discussed in the previous subsection.
The spectra are arranged in order of increasing distance from the centre of the 
galaxy. The positions of the the main emission line features are indicated in each
panel. The C IV $\lambda$5808 line is clearly detectable 
in the first two of the central spaxel spectra
and shows a broad component. The Wolf Rayet features are particularly strong in this
galaxy; spectral stacking may be a way to conduct systematic studies of Wolf-Rayet
signatures for larger samples of bulges in future work in order to study possible
variations in the upper IMF more systematically.

\section{Summary}

The goal of this paper has been to search for galaxy centers with clear
indications of unusual stellar populations. The motivation for this is the
finding that the stellar content of our own Galactic Center is surprisingly
exotic.  Two  inner disks of massive young stars are found that must have
formed very recently with an initial mass function that is flatter than
Salpeter at high stellar masses. These unusual stellar populations are only
found within the central parsec of our own Galaxy and would be undetectable
in external galaxies with current telescopes and instrumentation. The
hypothesis that is tested in this analysis is that there are populations of
bulges where similar phenomena extend over much larger scales, where the
total mass in young high mass stars is much larger, and where the impact of
these stars on the galaxy as they evolve could be quite dramatic.

Out of a sample of 668 face-on galaxies with stellar masses in the range
$10^{10}- 10^{11} M_{\odot}$, I identify 15 galaxies with young to
intermediate age inner stellar populations and where stellar population
gradients show unusual patterns in their central regions. In these
galaxies, the 4000 \AA\ break is either flat or rising towards the center
of the galaxy, indicating that the central regions  host evolved stars,
but the H$\alpha$ equivalent width also rises in the central regions,
indicating the presence of increasing amounts of strongly ionized gas. The
ionization parameter [OIII]/[OII] is typically low in these  galactic
centers, indicating that ionizing sources are stellar rather than AGN. Wolf
Rayet features characteristic of hot young stars are often found in the
spectra and these also get progressively stronger at smaller galactocentric
radii.

These outliers are compared to a control sample of galaxies of similar mass
with young inner stellar populations, but where the gradients in H$\alpha$
equivalent width and 4000 \AA\ break follow each other more closely. It is
demonstrated that central Wolf Rayet excesses are much more common in the
outliers, that the outliers have higher central stellar and ionized gas
metallicities, and that they are also much more frequently detected in the
radio compared to the control galaxies. I highlight one outlier where the
ionized gas is clearly being strongly perturbed and blown out either by
massive stars after they explode as supernovae, or by energy injection from
matter falling onto black holes.

Detailed quantitative comparisons with and inferences from stellar
population synthesis  models will be the subject of future work. I have
highlighted a number of the stumbling blocks with regard to accurate
population synthesis modelling that will first need to be overcome. It will
also be interesting to probe the structure and kinematics of the gas and
stellar distributions at the centers of the closest of these outliers with
high resolution spatially resolved observations, particularly at radio and
sub-millimetre wavelengths. It would also be very interesting to understand
the connection between the outlier population studied in this paper and
the nuclear star cluster phenomenon. Do the objects selected using our
methodology consititute the formation phase of the most massive of such
systems? The qualitative similarities  with the stellar properties 
of the nuclear star cluster in
our own Milky Way hints that there is much left to be understood about
the  role of very massive stars in the formation and evolution of all
galactic bulges.\\
\vspace {1cm}

\large
{\bf Acknowledgements}\\
\normalsize

I thank Reinhard Genzel, Eckhardt Sturm, Thorsten Naab
and Claudia Maraston for helpful
discussions about this work.

Funding for SDSS-IV has been provided by the Alfred
P. Sloan Foundation and Participating Institutions. Ad-
ditional funding towards SDSS-IV has been provided by
the US Department of Energy Onece of Science. SDSS-
IV acknowledges support and resources from the Centre
for High-Performance Computing at the University of
Utah. The SDSS web site is www.sdss.org.
SDSS-IV is managed by the Astrophysical Research
Consortium for the Participating Institutions of the
SDSS Collaboration including the Brazilian Participation
Group, the Carnegie Institution for Science,
Carnegie Mellon University, the Chilean Participation
Group, the French Participation Group, Harvard-
Smithsonian Center for Astrophysics, Instituto de
Astrofsica de Canarias, The Johns Hopkins University
sity, Kavli Institute for the Physics and Mathematics
of the Universe (IPMU)/University of Tokyo, Lawrence
Berkeley National Laboratory, Leibniz Institut fur 
Astrophysik Potsdam (AIP), Max-Planck-Institut f\"ur 
Astronomie (MPIA Heidelberg), Max-Planck-Institut f\"ur
Astrophysik (MPA Garching), Max-Planck-Institut f\"ur
Extraterrestrische Physik (MPE), National Astronom-
ical Observatory of China, New Mexico State University,
New York University, University of Notre Dame,
Observatario Nacional/MCTI, the Ohio State University,
Pennsylvania State University, Shanghai Astronomical
Observatory, United Kingdom Participation Group,
Universidad Nacional Autonoma de Mexico, University
of Arizona, University of Colorado Boulder, University
of Oxford, University of Portsmouth, University of Utah,
University of Virginia, University of Washington, 
University of Wisconsin, Vanderbilt University and Yale
University.

\vspace {1cm}
\noindent
{\bf Data availability:} The data is available upon reasonable request to the
corresponding author.


\end{document}